\documentclass[journal]{IEEEtran}
\usepackage{cite}
\usepackage{amsmath,amssymb,amsfonts,amsthm}
\usepackage{algorithm}
\usepackage{algpseudocode}
\usepackage{graphicx}
\usepackage{textcomp}
\usepackage{xcolor}
\usepackage{multirow}
\usepackage[dvipdfmx]{hyperref}
\usepackage{float}
\usepackage{acro} 
\usepackage{mfirstuc}

\DeclareAcronym{dl}{short=DL, long=deep learning}
\DeclareAcronym{ldpc}{short=LDPC, long=low-density parity check}
\DeclareAcronym{xr}{short=XR, long=extended reality}
\DeclareAcronym{v2x}{short=V2X, long=vehicle-to-everything}
\DeclareAcronym{embb}{short=eMBB, long=enhanced mobile broadband}
\DeclareAcronym{urllc}{short=URLLC, long=ultra-reliable low-latency communication}
\DeclareAcronym{mmtc}{short=mMTC, long=massive machine type communication}
\DeclareAcronym{kpi}{short=KPI, long=key performance indicator}
\DeclareAcronym{ml}{short=ML, long=machine learning}
\DeclareAcronym{gpu}{short=GPU, long=graphics processing unit}
\DeclareAcronym{cuda}{short=CUDA, long=compute unified device architecture}
\DeclareAcronym{fpga}{short=FPGA, long=field programmable gate array}
\DeclareAcronym{tpu}{short=TPU, long=tensor processing unit}
\DeclareAcronym{ae}{short=AE, long=autoencoder}
\DeclareAcronym{mu}{short=MU, long=multi-user}
\DeclareAcronym{mimo}{short=MIMO, long=multiple-input multiple-output}
\DeclareAcronym{3gpp}{short=3GPP, long=third generation partnership project}
\DeclareAcronym{dnn}{short=DNN, long=deep neural network}
\DeclareAcronym{mlp}{short=MLP, long=multi-layer perceptron}
\DeclareAcronym{relu}{short=ReLU, long=rectified linear unit}
\DeclareAcronym{bgd}{short=BGD, long=batch gradient descent}
\DeclareAcronym{sgd}{short=SGD, long=stochastic gradient descent}
\DeclareAcronym{mbsgd}{short=MBSGD, long=mini-batch stochastic gradient descent}
\DeclareAcronym{llm}{short=LLM, long=large language model}
\DeclareAcronym{rl}{short=RL, long=reinforcement learning}
\DeclareAcronym{mdp}{short=MDP, long=Markov decision process}
\DeclareAcronym{drl}{short=DRL, long=deep reinforcement learning}
\DeclareAcronym{dqn}{short=DQN, long=deep Q-network}
\DeclareAcronym{ber}{short=BER, long=bit error rate}
\DeclareAcronym{papr}{short=PAPR, long=peak-to-average power ratio}
\DeclareAcronym{ofdm}{short=OFDM, long=orthogonal frequency division multiplexing}
\DeclareAcronym{cnn}{short=CNN, long=convolutional neural network}
\DeclareAcronym{nlp}{short=NLP, long=natural language processing}
\DeclareAcronym{ilsvrc}{short=ILSVRC, long=ImageNet Large Scale Visual Recognition Challenge}
\DeclareAcronym{rnn}{short=RNN, long=reccurent neural network}
\DeclareAcronym{lstm}{short=LSTM, long=long short term memory}
\DeclareAcronym{gru}{short=GRU, long=gated recurrent unit}
\DeclareAcronym{gnn}{short=GNN, long=graph neural network}
\DeclareAcronym{mpnn}{short=MPNN, long=message passing neural network}
\DeclareAcronym{gpt}{short=GPT , long=generative pre-trained Transformer}
\DeclareAcronym{llama}{short=LLaMA, long=Large Language Model Meta AI}
\DeclareAcronym{palm}{short=PaLM, long=Pathways Language Model}
\DeclareAcronym{dm}{short=DM, long=diffusion model}
\DeclareAcronym{ddpm}{short=DDPM, long=denoising diffusion probabilistic model}
\DeclareAcronym{de}{short=DE, long=density evolution}
\DeclareAcronym{exit}{short=EXIT, long=extrinsic information transfer}
\DeclareAcronym{bp}{short=BP, long=belief propagation}
\DeclareAcronym{awgn}{short=AWGN, long=additive white Gaussian noise}
\DeclareAcronym{biawgn}{short=BI-AWGN, long=binary-input additive white Gaussian noise}
\DeclareAcronym{sc}{short=SC, long=successive cancellation}
\DeclareAcronym{vn}{short=VN, long=variable node}
\DeclareAcronym{cn}{short=CN, long=check node}
\DeclareAcronym{snr}{short=SNR, long=signal-to-noise ratio}
\DeclareAcronym{ga}{short=GA, long=Gaussian approximation}
\DeclareAcronym{rca}{short=RCA, long=reciprocal channel approximation}
\DeclareAcronym{crc}{short=CRC, long=cyclic redundancy check}
\DeclareAcronym{cascl}{short=CA-SCL, long=cyclic redundancy check-aided successive cancellation list}
\DeclareAcronym{bler}{short=BLER, long=block error rate}
\DeclareAcronym{pgd}{short=PGD, long=projected gradient descent}
\DeclareAcronym{pccmp}{short=PCCMP, long=polar-code-construction message-passing}
\DeclareAcronym{a2c}{short=A2C, long=advantage actor-critic}
\DeclareAcronym{isit}{short=ISIT, long=International Symposium on Information Theory}
\DeclareAcronym{pac}{short=PAC, long=polarization adjusted convolutional}
\DeclareAcronym{map}{short=MAP, long=maximum a posterior}
\DeclareAcronym{bigru}{short=Bi-GRU, long=bidirectional gated recurrent unit}
\DeclareAcronym{ecct}{short=ECCT, long=error correction code Transformer}
\DeclareAcronym{pcm}{short=PCM, long=parity check matrix}
\DeclareAcronym{bpsk}{short=BPSK, long=binary phase shift keying}
\DeclareAcronym{bch}{short=BCH, long=Bose–Chaudhuri–Hocquenghem}
\DeclareAcronym{siso}{short=SISO, long=soft-input soft-output}
\DeclareAcronym{mind}{short=MIND, long=model independent neural decoder}
\DeclareAcronym{osd}{short=OSD, long=ordered statistic decoding}
\DeclareAcronym{sttmram}{short=STT-MRAM, long=spin-torque transfer magnetic random access memory}
\DeclareAcronym{ms}{short=MS, long=min-sum}
\DeclareAcronym{nms}{short=NMS, long=normalized min-sum}
\DeclareAcronym{oms}{short=OMS, long=offset min-sum}
\DeclareAcronym{noms}{short=NOMS, long=normalized offset min-sum}
\DeclareAcronym{ams}{short=AMS, long=adjusted min-sum}
\DeclareAcronym{smms}{short=SMMS, long=single-minimum min-sum}
\DeclareAcronym{vwms}{short=VWMS, long=variable weight min-sum}
\DeclareAcronym{scms}{short=SCMS, long=self-corrected min-sum}
\DeclareAcronym{rbms}{short=RBMS, long=reliability-based min-sum}
\DeclareAcronym{svm}{short=SVM, long=support vector machine}
\DeclareAcronym{rlm}{short=RLM, long=regularized loss minimization}
\DeclareAcronym{lams}{short=LAMS, long=min-sum decoding with linear approximation}
\DeclareAcronym{pbldpc}{short=PB-LDPC, long=protograph-based low-density parity-check}
\DeclareAcronym{lut}{short=LUT, long=look-up table}
\DeclareAcronym{llr}{short=LLR, long=log-likelihood ratio}
\DeclareAcronym{hdd}{short=HDD, long=hard-decision decoder}
\DeclareAcronym{rrd}{short=RRD, long=random redundant decoding}
\DeclareAcronym{mrrd}{short=mRRD, long=modified random redundant decoding}
\DeclareAcronym{mbbp}{short=MBBP, long=multiple-bases belief-propagation}
\DeclareAcronym{ncrd}{short=NC-RD, long=node-classified redundant decoding}
\DeclareAcronym{hdpc}{short=HDPC, long=high-density parity-check}
\DeclareAcronym{lp}{short=LP, long=linear programming}
\DeclareAcronym{admm}{short=ADMM, long=alternating direction method of multipliers}
\DeclareAcronym{gldpc}{short=GLDPC, long=generalized low-density parity-check}
\DeclareAcronym{faid}{short=FAID, long=finite alphabet iterative decoder}
\DeclareAcronym{rqnn}{short=RQNN, long=recurrent quantized neural network}
\DeclareAcronym{ste}{short=STE, long=straight-through estimator}
\DeclareAcronym{tt}{short=TT, long=tensor-train}
\DeclareAcronym{tr}{short=TR, long=tensor-ring}
\DeclareAcronym{harq}{short=HARQ, long=hybrid automatic repeat request}
\DeclareAcronym{pnn}{short=PNN, long=partitioned neural network}
\DeclareAcronym{spc}{short=SPC, long=single parity check}
\DeclareAcronym{rc}{short=RC, long=repetition code}
\DeclareAcronym{nsc}{short=NSC, long=neural successive cancellation}
\DeclareAcronym{bf}{short=BF, long=bit flipping}
\DeclareAcronym{scl}{short=SCL, long=successive cancellation list}
\DeclareAcronym{scf}{short=SCF, long=successive cancellation flip}
\DeclareAcronym{sclf}{short=SCLF, long=successive cancellation list flip}
\DeclareAcronym{fscl}{short=FSCL, long=fast successive cancellation list}
\DeclareAcronym{fscf}{short=FSCF, long=fast successive cancellation flip}
\DeclareAcronym{fsclf}{short=FSCLF, long=fast successive cancellation list clip}
\DeclareAcronym{dscf}{short=DSCF, long=dynamic successive cancellation flip}
\DeclareAcronym{dsclf}{short=DSCLF, long=dynamic successive cancellation list flip}
\DeclareAcronym{dnc}{short=DNC, long=differentiable neural computer} 
\DeclareAcronym{sp}{short=SP, long=shifted-pruning} 
\DeclareAcronym{itd}{short=ITD, long=iterative threshold decoding}
\DeclareAcronym{hornn}{short=HORNN, long=high-order recurrent neural network} 
\DeclareAcronym{rsc}{short=RSC, long=recursive systematic convolutional} 
\DeclareAcronym{bcjr}{short=BCJR, long=Bahl-Cocke-Jelinek-Raviv} 
\DeclareAcronym{rm}{short=RM, long=Reed-Muller}
\DeclareAcronym{xai}{short=XAI, long=explainable AI}
\DeclareAcronym{qml}{short=QML, long=quantum machine learning}
\DeclareAcronym{nisq}{short=NISQ, long=noisy intermediate-scale quantum}
\DeclareAcronym{vqe}{short=VQE, long=variational quantum eigensolver}
\DeclareAcronym{qaoa}{short=QAOA, long=quantum approximate optimization algorithm}
\DeclareAcronym{gan}{short=GAN, long=generative adversarial network}
\DeclareAcronym{isi}{short=ISI, long=inter-symbol interference}
\DeclareAcronym{biso}{short=BISO, long=binary-input symmetric-output}

\makeatletter
\let\MYcaption\@makecaption
\makeatother
\usepackage{subcaption}
\makeatletter
\let\@makecaption\MYcaption
\makeatother

\def\BibTeX{{\rm B\kern-.05em{\sc i\kern-.025em b}\kern-.08em
    T\kern-.1667em\lower.7ex\hbox{E}\kern-.125emX}}
    
\begin{document}

\title{
Recent Advances in Deep Learning \\for Channel Coding: A Survey
}

\author{
  Toshiki Matsumine,~\IEEEmembership{Member,~IEEE},
  and Hideki Ochiai,~\IEEEmembership{Fellow,~IEEE}
  \thanks{T. Matsumine is with the Institute of Advanced Sciences, Yokohama National University, Yokohama, Japan (e-mail: matsumine-toshiki-mh@ynu.ac.jp)}
  \thanks{H. Ochiai is with the Graduate School of Engineering, Osaka University, Osaka, Japan (e-mail: ochiai@comm.eng.osaka-u.ac.jp)}
}

\maketitle
\IEEEpeerreviewmaketitle

\begin{abstract}
This paper provides a comprehensive survey on recent advances in deep learning (DL) techniques for the channel coding problems. Inspired by the recent successes of DL in a variety of research domains, its applications to the physical layer technologies have been extensively studied in recent years, and are expected to be a potential breakthrough in supporting the emerging use cases of the next generation wireless communication systems such as 6G. In this paper, we focus exclusively on the channel coding problems and review existing approaches that incorporate advanced DL techniques into code design and channel decoding. After briefly introducing the background of recent DL techniques, we categorize and summarize a variety of approaches, including model-free and mode-based DL, for the design and decoding of modern error-correcting codes, such as low-density parity check (LDPC) codes and polar codes, to  highlight their potential advantages and challenges. Finally, the paper concludes with a discussion of open issues and future research directions in channel coding.
\end{abstract}


\section{Introduction}
\label{sec:intro}
Channel coding is a well-established area of research with a long history dating back to the Shannon's theory \cite{shannon1948mathematical} where he introduced the \emph{Shannon limit} as the maximum rate at which information can be transmitted over a given communication channel. 
Subsequently, researchers have made tremendous efforts to develop a practical coding scheme that approaches the Shannon limit at a realistic implementation cost \cite{costello2007channel}.
The notable successes in coding theory include the invention of modern capacity-approaching codes such as turbo codes \cite{berrou1993near}, \ac{ldpc} codes \cite{gallager1962low}, and polar codes \cite{arikan2009channel}. These coding techniques have contributed significantly to various communication systems, such as wired and wireless communications, as well as storage systems, from the viewpoint of improving reliability and energy efficiency.

Due to the emerging wireless applications, including \ac{xr} for telemedicine, tactile Internet, \ac{v2x}, and wireless data centers, the next generation wireless communication systems impose unprecedentedly diverse and stringent requirements for, e.g., ultra-high data rate, ultra-low latency, and high energy efficiency \cite{yang20196g,zhang20196g,huang2019survey,saad2019vision,letaief2019roadmap,chowdhury20206g,dogra2020survey,gui20206g,kato2020ten,giordani2020toward,rajatheva2020white,chen2020vision,dang2020should,akyildiz20206g,khan20206g,yaacoub2020key,viswanathan2020communications,tariq2020speculative,tataria20216g,nguyen20216g,jiang2021road,alsabah20216g,de2021convergent,de2021survey,guo2021enabling,matthaiou2021road,noor20226g,wang2023road,chafii2023twelve,you2024next}. In 5G, the main use cases are \ac{embb}, \ac{urllc}, and \ac{mmtc}, and for each case, the system requirement is specified in terms of a single \ac{kpi}, such as throughput, latency, reliability, and energy efficiency. On the other hand, due to the diversity of applications, many use cases in the next wireless communications such as 6G will require trade-offs among different \acp{kpi}, which poses a new challenge for the design of physical layer techniques \cite{wang2023road,geiselhart20236g,zhang2023channel}.

Traditionally, the design of coding schemes has been based on mathematical models and expert knowledge, such as coding theory and information theory. Although this approach has contributed significantly to the recent progress in practical channel coding, it also has limitations. Specifically, it relies on mathematical models that do not fully capture real-world environments, and thus there is always a mismatch between the model for which we design systems and the actual environment to which they are applied. In addition, in the next generation communications, the system design problem will become increasingly complex due to demanding requirements and therefore will not be mathematically tractable in most cases. To address these issues, \emph{data-driven} approaches to communication system design based on \ac{ml} techniques have emerged as a new paradigm that supports or replaces the conventional system design based on mathematical models. 

In particular, inspired by the recent successes of \ac{dl} technologies in broad research areas, their applications in communication systems have been extensively studied. This DL trend has been accelerated by the development of dedicated DL frameworks, such as Tensorflow \cite{abadi2016tensorflow} and Pytorch  \cite{collobert2011torch7}, which makes it easier for researchers to implementate their DL algorithms. Furthermore, most of them are built with \ac{gpu} acceleration provided by the NVIDIA \ac{cuda} Deep Neural Network library (cuDNN), which significantly speeds up DL training due to its ability to perform parallel computations and high memory bandwidth. In addition, various DL processors such as \acp{fpga} and \ac{tpu} have been explored in the literature \cite{jouppi2017datacenter,qasaimeh2019comparing}, enabling efficient hardware implementations of DL-based communications and networking in beyond 5G and 6G.

\begin{table*}[h]
    \centering
    \caption{Survey papers related to our work.}
    \begin{tabular}{|p{0.5cm}|p{2.5cm}|p{12cm}|} \hline
        Year & Reference & Contents related to channel coding. \\ \hline

        \multirow{1}{*}{2017} & Wang et al. \cite{wang2017deep} & 
        Review of early works on DL-based decoding. \\

        \hline
        
        \multirow{3}{*}{2019} & Zhang et al. \cite{zhang2019deep} & 
        Brief introduction of DL-based channel decoding. \\ \cline{2-3}
                
                              & Gunduz et al. \cite{gunduz2019machine} & 
        Brief review of DL-aided decoding. \\ \cline{2-3}
               
                              & Balatsoukas et al. \cite{balatsoukas2019deep} & 
        Review of deep unfolding for channel decoding. \\  
        
        \hline
        
        \multirow{2}{*}{2020} & Samad et al. \cite{samad2020white} & 
        Addressing DL-based channel decoding. \\ \cline{2-3}
                        
                              & Zhang et al. \cite{zhang2020artificial} & 
        Review of DL-based decoding and code construction. \\

        \hline
                        
        \multirow{1}{*}{2021} & Ly et al. \cite{ly2021review} & 
        Review of DL applications for LDPC code identification, decoding. \\
        
        \hline
                
        \multirow{2}{*}{2023} & Mao et al. \cite{mao2023deep} & 
         Briefly addressing DL-aided decoder and genetic algorithm for code construction. \\ \cline{2-3}

                              & Akrout et al. \cite{akrout2023domain} & 
         Domain generalization \cite{zhou2022domain,wang2022generalizing} in the channel decoding problem. \\

         \hline
                 
        \multirow{2}{*}{2024} & Ye et al. \cite{ye2024artificial} & 
        Review of DL-based decoding for turbo, LDPC, and polar codes. \\ \cline{2-3}

                              & Rowshan et al. \cite{rowshan2024channel} &
        Comprehensive survey on channel coding with brief introduction of DL for channel decoding.
          \\

        \hline
    \end{tabular}
  \label{tab:survey}
\end{table*}

\subsection{DL Applications to The Physical Layer}
In recent years, DL has been successfully applied to the physical layer of communication systems
\footnote{We note that ML techniques for the physical layer have been studied sporadically for many years, e.g., in \cite{wang1996artificial,hamalainen1999recurrent,ibnkahla2000applications}. However, the applications of DL are rather new, which started to flourish around 2016.} \cite{o2017introduction,wang2017deep,simeone2018very,dorner2018deep,gunduz2019machine,huang2019deep,he2019model,kim2019toward,varasteh2020learning,samad2020white,restuccia2020deep,zhang2020artificial,ly2021review,matthaiou2021road,zhang2021deep,ozpoyraz2022deep,lu2023semantics,mao2023deep,ohtsuki2023machine,shi2023machine,akrout2023domain,van2024generative,islam2024deep,hoang2024physical}.
One of the seminal works is \cite{o2017introduction}, where the authors introduced the concept of end-to-end learning for communication systems, which is considered as \ac{ae}. The authors also introduced other DL applications, such as modulation classification and radio transformer networks.
Later on, many papers discussed potential applications of DL to communication problems, such as channel decoding, signal detection, channel modeling, and \ac{mimo} signal detection \cite{wang2017deep,simeone2018very,dorner2018deep,gunduz2019machine,huang2019deep,he2019model,kim2019toward}.

One of the notable advances in DL for the physical layer is the development of an open source Python library, called \emph{Sionna}, which was released by NVIDIA in 2022 \cite{hoydis2022sionna}. It supports link-level simulation of \ac{mu}-MIMO systems with 5G-compliant channel codes, the \ac{3gpp} channel models, channel estimation, and so forth. Each building block is implemented using TensorFlow and allows for gradient-based optimization via backpropagation. Sionna alo has a native NVIDIA GPU support.

As the number of publications related to DL-based approaches for physical layer technologies has been increasing almost exponentially, it is important to classify and summarize them to highlight the current state and challenges. However, despite its importance, only a handful of survey papers have been dedicated to the channel coding problems so far.

\subsection{Our Scope and Related Works}
Existing papers on DL for channel coding may be classified into the following categories:
\begin{enumerate}
    \item DL-based code design,
    \item DL-based channel decoding,
    \item End-to-end learning for communication systems.
\end{enumerate}
In DL-based code design, the optimization of code parameters, such as the degree distribution of LDPC codes and the locations of frozen bits in polar codes, is performed using DL techniques.
Channel decoding is a popular DL application as the decoding problem is essentially the classification problem which is what DL is good at. This approach utilizes a \ac{dnn} to replace or augment a conventional channel decoder with the purpose of improving the error correction performance or reducing the complexity and/or latency.
End-to-end learning of communication systems is another popular application, where a transmitter-receiver pair is often completely replaced by ``black-box'' DNNs and trained over a differential channel model in an end-to-end fashion. In this approach, not only a channel encoder-decoder pair can be trained, but also other physical layer components such as source encoder-decoder and symbol mapper-demapper.

Although end-to-end learning has been extensively studied in the literature, and this approach would be particularly promising for a new paradigm of semantic communication \cite{xie2021deep,qin2021semantic,xu2022full,yang2022semantic,luo2022semantic,gunduz2022beyond,lu2023semantics,getu2023making,li2023towards,wheeler2023engineering,shi2023machine,islam2024deep}, the complete replacement of transceivers by DNNs poses several challenges in practice and does not seem feasible at this time. Therefore, this paper focuses on DL-based approaches that may be applicable to existing systems with appropriate modifications. In particular, we consider code design using offline DL techniques and DL-based channel decoding to replace or support conventional decoders\footnote{Although we mainly focus on \emph{classical} error-correcting codes in this paper, DL-based code design and decoding has also been extensively studied in the realm of \emph{quantum} error-correcting codes, e.g., in \cite{torlai2017neural,varsamopoulos2017decoding,varsamopoulos2019comparing,nautrup2019optimizing,carleo2019machine,iolius2023decoding,krenn2023artificial}.}.

In Table~\ref{tab:survey}, we summarize existing survey papers related to our scope in this work. We emphasize that most of the existing survey papers cover a wide range of DL applications in the physical layer, rather than dealing with DL-assisted channel coding in a comprehensive manner. Nevertheless, the paper most closely related to our work would be \cite{zhang2020artificial}, in which the authors discussed a wide range of DL applications in the physical layer, including the channel coding problems such as channel decoding and code construction. However, since its publication in 2020, a significant number of new techniques have been proposed. By focusing solely on the channel coding problems, we attempt to provide a comprehensive survey including the state-of-the-art techniques.

\begin{figure*}[t]
\centering
\includegraphics[width=0.99\linewidth]{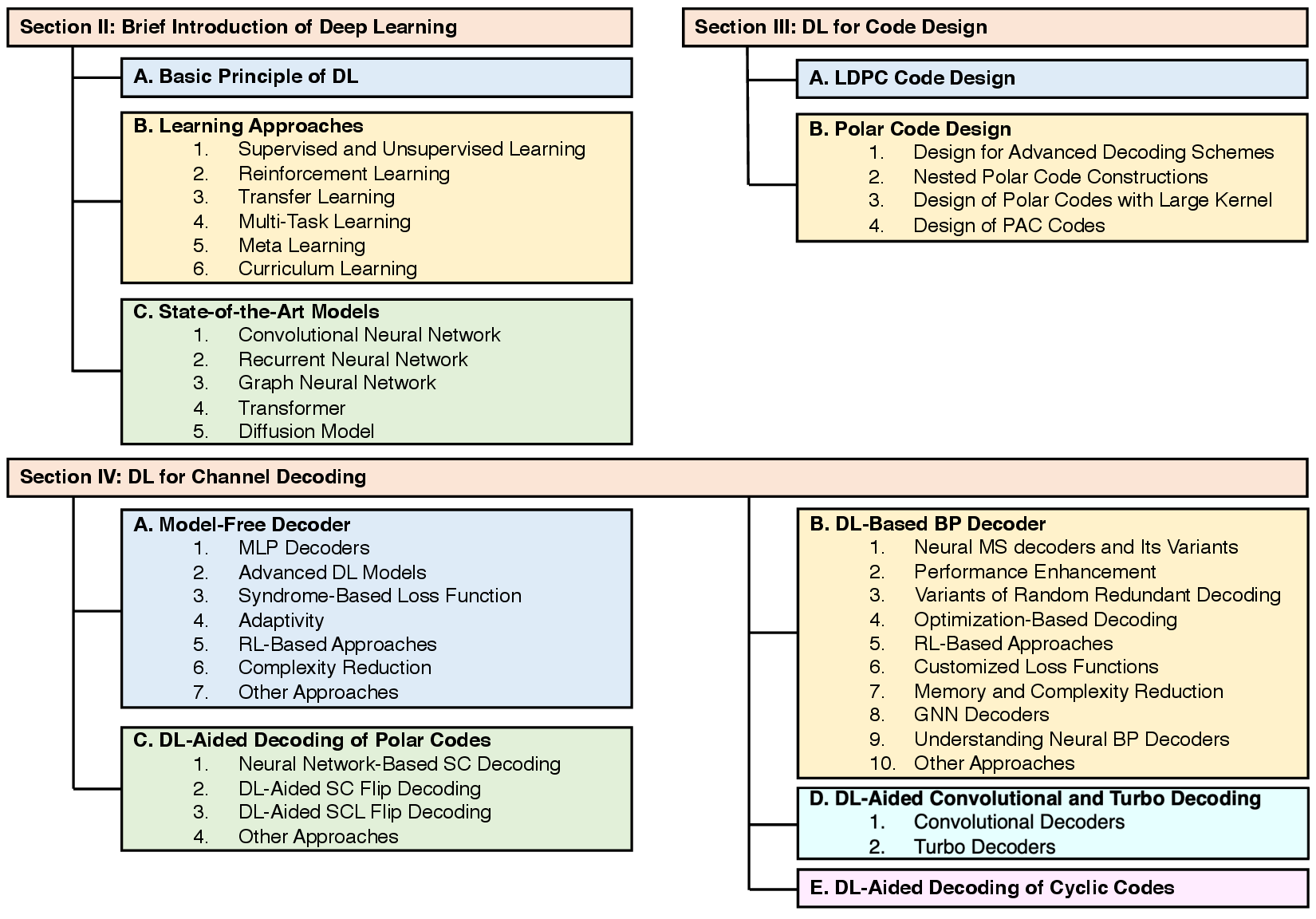}
\caption{Organization of this survey paper.}
\label{fig:section} 
\end{figure*}

The overall organization of this paper is visualized in Fig.~\ref{fig:section}.
In Section~\ref{sec:dl}, we briefly introduce the basics of DL techniques and the state-of-the-art models to facilitate the understanding of our survey. In Section~\ref{sec:design}, we review DL-based design of LDPC and polar codes. Then, in Section~\ref{sec:decoding}, we consider various DL approaches to the channel decoding problem. Finally, we conclude this paper by discussing the challenges and future directions in Section~\ref{sec:conclusion} to stimulate further research.
We provide a list of abbreviations that we use after Section~\ref{sec:conclusion}.

\section{Brief Introduction of Deep Learning}
\label{sec:dl}
In this section, we briefly review the basics of DL technologies, starting with neural networks and their optimization. We then introduce the state-of-the-art training and DL models. For details on the theory of general DL techniques, please refer, e.g., to 
\cite{mcculloch1943logical,rumelhart1986learning,schmidhuber2015deep,lecun2015deep,liu2017survey,deng2014deep,pouyanfar2018survey,alom2019state,dong2021survey,dargan2020survey}.

\subsection{Basic Principle of DL}
DL is a subfield of ML that uses DNNs with multiple hidden layers between input and output layers.
Among various neural network structures, \ac{mlp} is a class of fully connected feedforward neural networks that consist of at least one hidden layer in addition to input and output layers \cite{rumelhart1985learning}. 

\begin{figure}[t]
\centering
\includegraphics[width=0.7\linewidth]{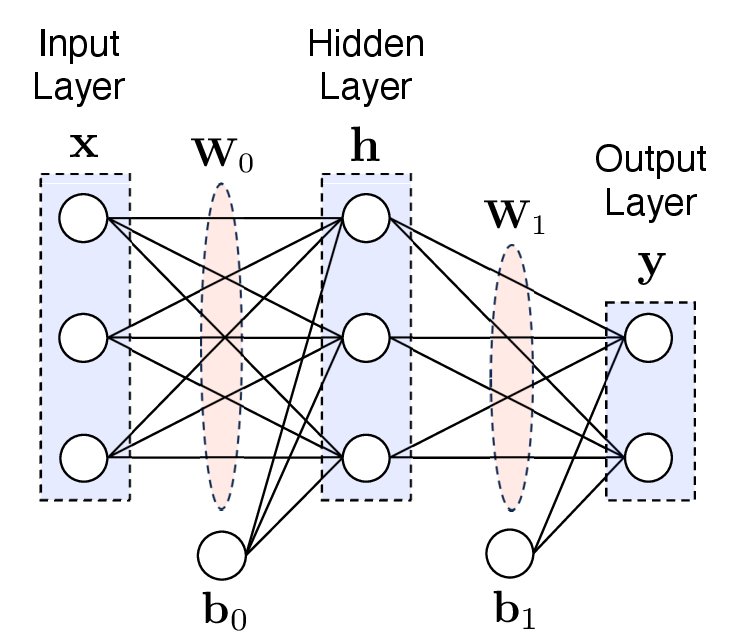}
\caption{An example of MLP with single hidden layer.}
\label{fig:mlp} 
\end{figure}
An example of a single hidden layer MLP is shown in Fig.~\ref{fig:mlp}, where the input vector $\mathbf{x} \in \mathbb{R}^3$ is mapped to the output vector $\mathbf{y} \in \mathbb{R}^2$ by applying a series of affine transformations and nonlinear activation functions as
\begin{align}
\mathbf{y} &= \phi_1(\mathbf{W}_1 \mathbf{h} + \mathbf{b}_1) \\   
           &= \phi_1(\mathbf{W}_1 \phi_0(\mathbf{W}_0 \mathbf{x} + \mathbf{b}_0) + \mathbf{b}_1),
\end{align}
where $\mathbf{W}_0 \in \mathbb{R}^{3 \times 3}$ and $\mathbf{W}_1 \in \mathbb{R}^{3 \times 2}$ are weight matrices, $\mathbf{b}_0 \in \mathbb{R}^{3}$ and $\mathbf{b}_1 \in \mathbb{R}^{2}$ are bias terms, and $\phi_i(\cdot)$ with $i \in \{0, 1\}$ denotes the element-wise application of a nonlinear activation function.

The nonlinear activation function allows the neural network to approximate highly complex functions, and the choice of activation functions has a significant impact on the resulting performance.
Although there are a number of activation functions \cite{dubey2022activation}, one of the most widely used modern activation functions is \ac{relu} \cite{nair2010rectified} and its variants such as Leaky ReLU \cite{maas2013rectifier} and parametric ReLU \cite{he2015delving}.

As for the optimization of the DNN parameters, i.e., $\Theta \triangleq \{\mathbf{W}_0, \mathbf{W}_1, \mathbf{b}_0, \mathbf{b}_1\}$ in our example, the most common approach is gradient descent, which is a first-order iterative algorithm for finding a local minimum of a differentiable function. The basic idea of gradient descent is to update the parameters in the opposite direction of the gradient of the differentiable loss function that we wish to minimize. Letting $f(\Theta)$ denote the loss function that is differentiable with respect to the parameter set $\Theta$, in the $i$-th iteration of gradient descent, the trainable parameters are updated as
\begin{align}
    \Theta_{i+1} = \Theta_{i} - \eta \nabla f(\Theta_{i}),
    \label{eq:sgd}
\end{align}
where $\eta \in \mathbb{R}_+$ is a learning rate that determines the size of the steps taken in the direction of steepest descent.

Gradient descent algorithms can be classified according to the amount of data used to compute the gradient, namely, \ac{bgd}, \ac{sgd}, and \ac{mbsgd}. BGD calculates the gradients for the entire training dataset, while SGD performs a parameter update (and thus gradient calculation) for each training data sample. Meanwhile, MBSGD substitutes small data batches for single samples in SGD, thereby reducing the variance of the parameter updates, which can lead to more stable convergence. Furthermore, the gradient computation in MBSGD can be efficiently performed by DL libraries. For these reasons, MBSGD is one of the most popular stochastic optimization methods for training DNNs.

However, the standard MBSGD does not necessarily guarantee good convergence, and many improvements have been proposed that adaptively control the learning rate during training.
These approaches include Momentum \cite{sutskever2013importance}, Adagrad \cite{duchi2011adaptive}, Adadelta \cite{zeiler2012adadelta}, RMSprop \footnote{Originally proposed in \url{http://www.cs.toronto.edu/~tijmen/csc321/slides/lecture_slides_lec6.pdf}.}, Nadam \cite{dozat2016incorporating}, and Adam \cite{kingma2014adam}.
Implementations of these optimizers are available in DL frameworks such as Tensorflow \cite{abadi2016tensorflow} and Pytorch \cite{collobert2011torch7}. For more details on SGD algorithms, see \cite{ruder2016overview}.

\subsection{Learning Approaches}
There are many training methods for ML techniques.
In the following, we introduce some of the major approaches that are often applied to the design of communication systems.

\subsubsection{Supervised and Unsupervised Learning}
\label{sec:sl}
Supervised learning trains algorithms based on labeled datasets consisting of pairs of inputs and corresponding correct outputs, i.e., ground truth. The goal is to analyze patterns from a large dataset and predict outcomes for new data. Supervised learning is commonly used for tasks such as classification and regression.
Since the channel decoding problem can be seen as a type of classification, the simplest approach would be to train a DL-based channel decoder, where a DNN is trained to estimate a transmitted codeword by minimizing the error between the correct and estimated codewords.

Unsupervised learning is another type of ML algorithm that learns patterns from data without human supervision. Self-supervised learning, often used to train \acp{llm}, can also be considered unsupervised learning in the sense that it uses the data itself to generate supervising signals, rather than relying on human supervision. There is also an approach called semi-supervised learning, which combines both supervised learning and unsupervised learning, i.e., it uses both labeled and unlabeled data. Although these approaches are particularly suitable for the practical scenario where a sufficiently large labeled dataset is not available, their applications to channel decoding are rather new topics to be investigated.

\subsubsection{Reinforcement Learning} 
\begin{figure}[t]
\centering
\includegraphics[width=0.8\linewidth]{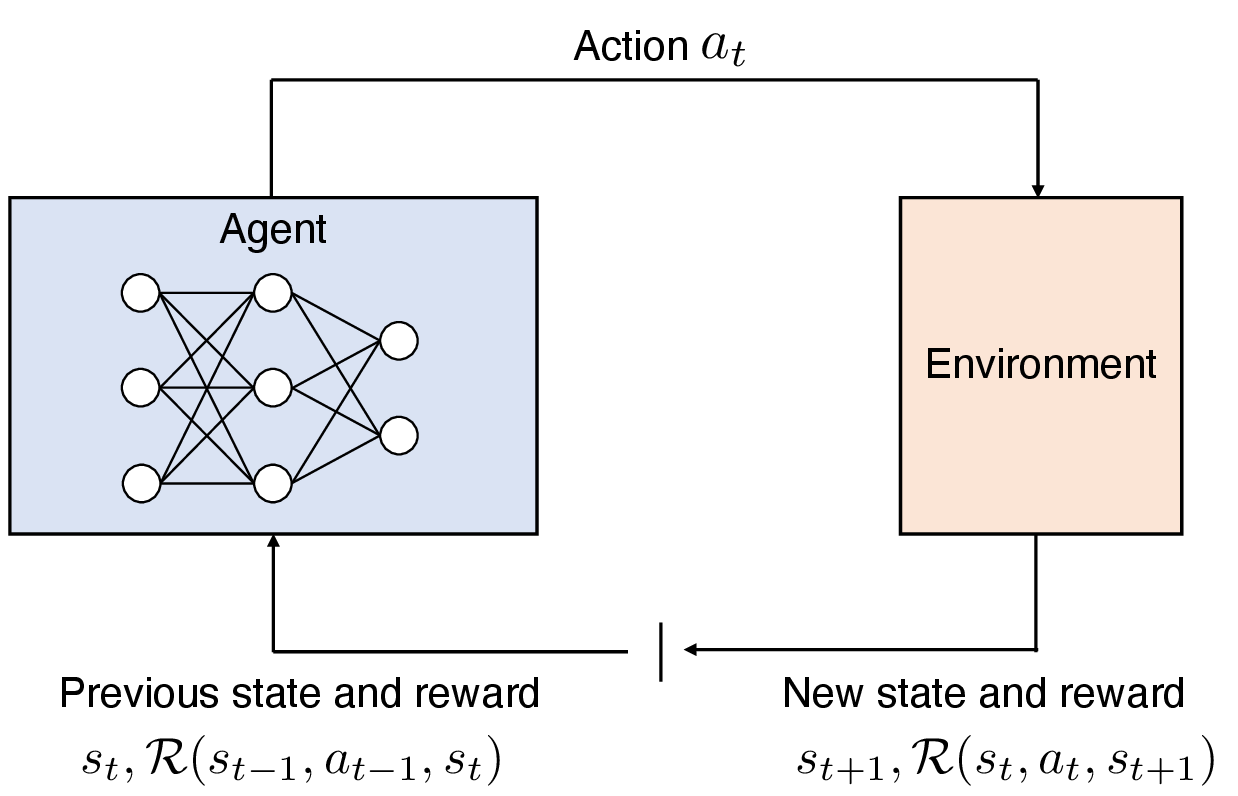}
\caption{A typical reinforcement learning framework.}
\label{fig:rl} 
\end{figure}

\Ac{rl} is an experience-driven autonomous learning framework where an intelligent agent learns to take actions in a dynamic environment in order to maximize the cumulative reward \cite{sutton2018reinforcement}.
RL is typically modeled as \ac{mdp} which consists of 
\begin{itemize}
    \item a set of environment and agent states $\mathcal{S}$
    \item a set of actions $\mathcal{A}$
    \item the transition probability from state $s_t$ to state $s_{t+1}$ upon action $a_t$ at time $t$, denoted by $\mathcal{P}(s_{t+1}|s_t, a_t)$, and the corresponding reward function $\mathcal{R}(s_t, a_t, s_{t+1})$.
\end{itemize}
The process is shown in Fig.~\ref{fig:rl}.
At each time step $t$, the agent observes a state $s_t \in \mathcal{S}$ and takes an action $a_t \in \mathcal{A}$, following a policy $\pi(a_t|s_t)$. Then the agent receives a scalar reward $r_t$, and transitions to the next state $s_{t+1}$, according to the reward function $\mathcal{R}(s_t, a_t, s_{t+1})$ and state transition probability $\mathcal{P}(s_{t+1}|s_t, a_t)$, respectively. This process continues until the agent reaches a terminal state. The return is the discounted, accumulated reward with the discount factor $\gamma \in (0, 1]$. The agent aims to maximize the expectation of such long term return from each state. 

In many practical problems, the states of the MDP are high-dimensional and difficult to solve with traditional RL algorithms. On the other hand, thanks to their powerful function approximation properties, the use of DNNs to approximate the optimal policy and/or the optimal value functions in RL provides an efficient way to overcome these problems. 
This approach, called \ac{drl}, has achieved remarkable results in a variety of research areas.
In particular, \ac{dqn} \cite{mnih2015human}, where a DNN model is built to approximate the Q function (the value of an action in a given state), showed impressive results in Atari \cite{mnih2015human}. More details on DRL can be found, for example, in \cite{arulkumaran2017deep,li2017deep,li2023deep,moerland2023model}.

\subsubsection{Transfer Learning}
\label{sec:tl}
\begin{figure}[t]
\centering
\includegraphics[width=0.8\linewidth]{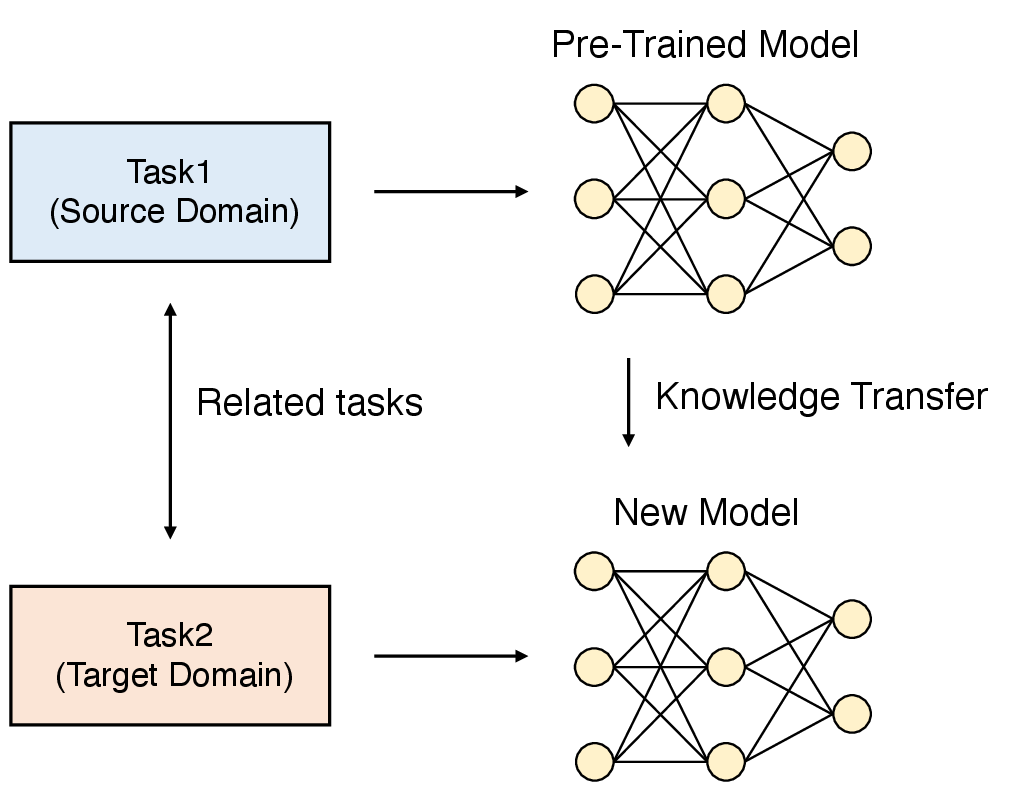}
\caption{The concept of transfer learning.}
\label{fig:transfer} 
\end{figure}

Transfer learning is a technique for transferring knowledge learned from a task in a source domain to imp performance on a related task in a target domain \cite{pan2009survey,weiss2016survey,tan2018survey,zhuang2020comprehensive,wang2021transfer}.
Transfer learning addresses the problem of insufficient labeled training data by transferring the knowledge across task domains. This concept is illustrated in Fig.~\ref{fig:transfer}. Note, however, that the transferred knowledge may be worthless if there is little or even nothing in common between the source and target domains.

In the channel coding problems, transfer learning can be useful to adapt DL-based code design and decoding trained for a certain channel model and code parameters to new channel models or parameters. For example, we usually train a DNN for design or decoding assuming a certain code rate, and adapting to a new code rate requires re-training of the DNN, which is time consuming and computationally expensive. Since codes derived from a single mother code by rate matching, i.e., puncturing and shortening, may have many similarities, transfer learning could be used to significantly reduce the computational burden of re-training or to improve the performance with the new parameter.

\subsubsection{Multi-Task Learning}
\label{sec:mtl}
In multi-task learning, a set of multiple tasks is solved jointly, sharing an inductive bias among them \cite{crawshaw2020multi,zhang2021survey}. Multi-task learning is inherently a multi-objective problem because different tasks may conflict, requiring a trade-off. A common compromise is to optimize a proxy objective that minimizes a weighted linear combination of per-task losses. Since this joint representation must capture useful features across all tasks, multi-task learning can hinder individual task performance if the different tasks seek conflicting representations, i.e., the gradients of different tasks point in opposing directions or differ significantly in magnitude. This phenomenon is commonly known as negative transfer.  

There are some similarities between transfer learning and multi-task learning. Both aim to improve learners' performance by knowledge transfer. On the other hand, the main difference is that the former transfers the knowledge contained in the related domains, while the latter transfers the knowledge by learning some related tasks simultaneously. In other words, multi-task learning pays equal attention to each task, while transfer learning pays more attention to the target task than to the source task. 

Similar to transfer learning, multi-task learning could be used to efficiently support multiple different code parameters. Furthermore, multi-task learning can be used for multi-objective optimization, which is common in many communication system design problems. Application examples of multi-task learning include AE-based constellation design that attempts to jointly minimize \ac{ber} and \ac{papr} for \ac{ofdm} systems \cite{kim2017novel}.

\subsubsection{Meta-Learning}
Unlike traditional learning approaches that attempt to solve tasks from scratch with a fixed algorithm, meta-learning aims to learn the learning algorithm itself by learning from previous experience or tasks \cite{vanschoren2018meta,hospedales2021meta}. This \emph{learning-to-learn} framework can lead to several benefits, such as improved data and computational efficiency.

In a meta-learning framework, there are two types of data, a larger data set of examples from related tasks (meta-training data) and a small training data set for a new task (meta-testing data). Standard meta-learning consists of two phases, 1) meta-training where a set of hyperparameters is optimized given the meta-training data set, and 2) meta-testing where model parameters, which are initialized with the meta-trained hyperparameters, are optimized using the meta-testing data. Thus, the meta-training phase aims to optimize hyperparameters that allow efficient training on a new, \emph{a priori} unknown, target task in the meta-testing phase.

Meta-learning could naturally be applied to adaptive decoder design, where the decoder parameters are initialized by meta-training and then optimized based on meta-testing to adapt to a new channel. In addition, the concept can be applied to a wide range of problems in the physical layer, such as signal demodulation, joint transmitter and receiver optimization via end-to-end learning, channel prediction, and so forth, as reviewed in \cite{chen2023learning}.

\subsubsection{Curriculum Learning}
Curriculum learning, originally proposed in \cite{bengio2009curriculum}, is a training strategy that trains a machine learning model from easier data to harder data, imitating human learning \cite{wang2021survey,soviany2022curriculum}. The basic idea is to “start small” \cite{elman1993learning}, i.e., to train the ML model with easier data subsets, and then gradually increase the difficulty level of the data until the entire training dataset is used. Curriculum learning can be seen as a special form of the continuation method which is a general strategy for global optimization of non-convex functions \cite{bengio2009curriculum}.

As the idea of curriculum learning serves as a general training strategy, it has been exploited in a considerable range of applications. In contrast to the standard training approach based on random data shuffling, curriculum learning can provide performance improvements with faster training convergence speed. In curriculum learning, the design of a curriculum strategy, i.e., a sequence of training criteria, plays a key role.

Curriculum learning can be beneficial for designing communication systems including channel coding schemes. For example, designing and decoding long codes is generally a more difficult task than designing and decoding short codes. As such, instead of learning to design or decode long codes directly, curriculum learning strategies that start from training with short codes and then gradually increase the code length have the potential to achieve not only faster training convergence, but also better performance.

\subsection{State-of-the-Art Models}
Finding an appropriate architecture is important when applying DL to the problems in communication systems. In the following, we briefly review the representative DL models that are useful for such problems. 

\subsubsection{Convolutional Neural Network}
\label{sec:cnn}
\begin{figure}[t]
\centering
\includegraphics[width=0.95\linewidth]{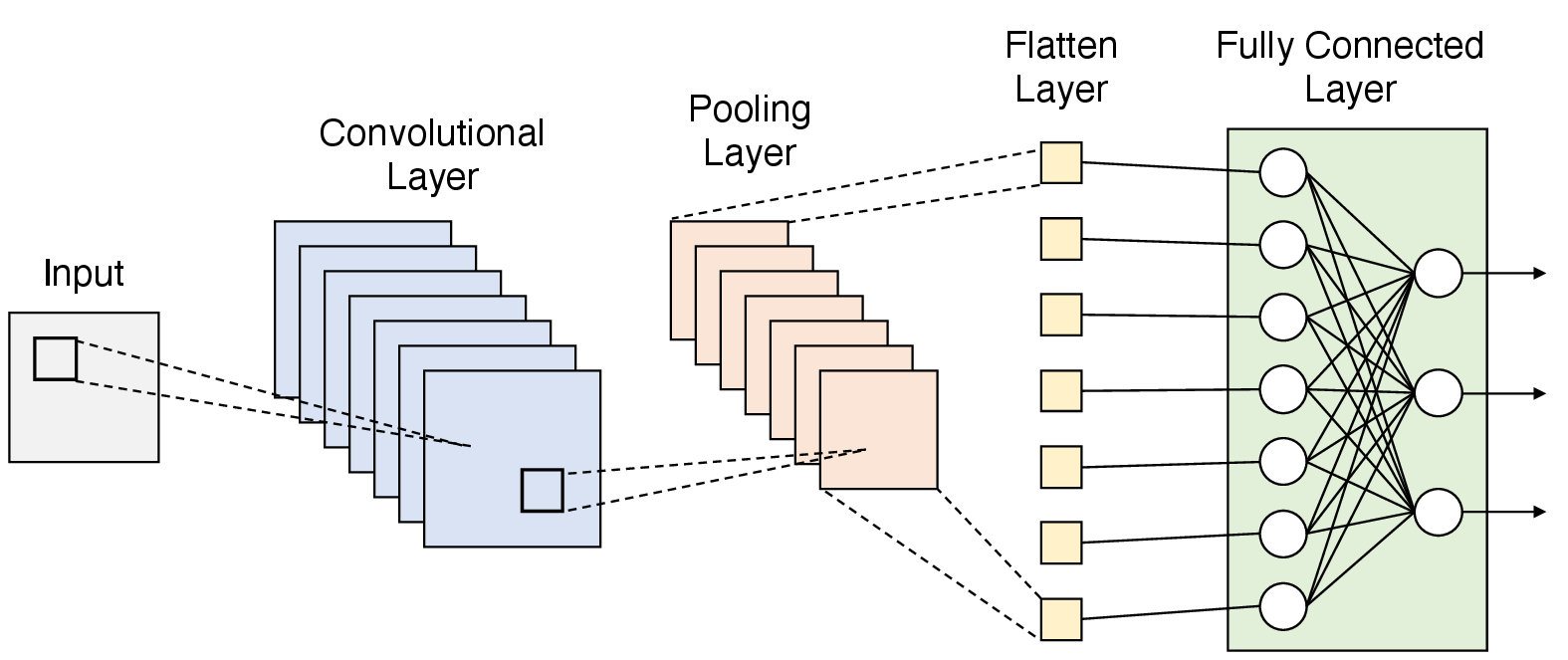}
\caption{A simple CNN architecture.}
\label{fig:cnn} 
\end{figure}

\Acp{cnn} have made tremendous success in a vast research field including computer vision and \ac{nlp} \cite{gu2018recent,li2021survey,krichen2023convolutional}. A CNN is a type of feed-forward neural network with a convolutional layer that learns features by applying filters (or kernels) to data. A simple example of a CNN architecture is illustrated in Fig.~\ref{fig:cnn}. The CNN has fewer connections and parameters than the MLP since each neuron in a convolutional layer receives input only from a restricted area of the previous layer. This restricted area is called the receptive field of the neuron, and in the case of a fully connected layer, the receptive field corresponds to the entire previous layer.

CNNs have been particularly successful in image recognition tasks, achieving state-of-the-art results on several benchmarks such as the \ac{ilsvrc}. Representative CNN models include AlexNet \cite{krizhevsky2012imagenet}, VGGnet \cite{simonyan2014very}, Inception (GoogLeNet) \cite{szegedy2015going}, ResNet \cite{he2016deep}, and DenseNet \cite{huang2017densely}. Their success is due to their ability to capture spatial features and patterns by using a hierarchical architecture of layers that perform convolution operations and extract features at different levels of abstraction \cite{chen2016deep}.

As for their applications in channel coding, they are often used for decoding, and have been demonstrated to be more efficient than standard MLP.
Other popular applications of CNN in the physical layer include channel estimation in time-frequency domain for OFDM systems \cite{soltani2019deep} and fully CNN receiver that replaces the conventional channel estimator, equalizer, and demapper \cite{honkala2021deeprx}.

\subsubsection{Recurrent Neural Network}
\label{sec:rnn}
Unlike unidirectional feedforward neural networks, a \ac{rnn} is a bi-directional artificial neural network that is capable of learning long-term dependencies from sequential data \cite{rumelhart1986learning,sutskever2014sequence}.
Due to their ability to use internal state (memory) to process arbitrary input sequences, they are particularly suited for processing time-series data such as speech recognition.
On the other hand, due to the recurrent connections, classical RNNs have the gradient vanishing and exploding problems, i.e., the long-term gradients may not converge and approach zero or infinity during backpropagation.

\begin{figure}[t]
     \centering
     \begin{subfigure}[b]{1.0\linewidth}
         \centering
         \includegraphics[width=0.4\linewidth]{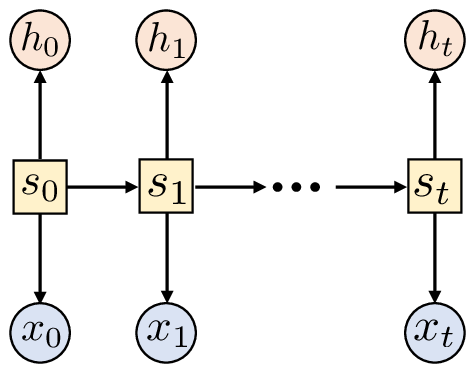}
         \caption{RNN.}
         \label{fig:rnn}
     \end{subfigure}
     \\ \vspace{3mm}
     \begin{subfigure}[b]{1.0\linewidth}
         \centering
         \includegraphics[width=0.7\linewidth]{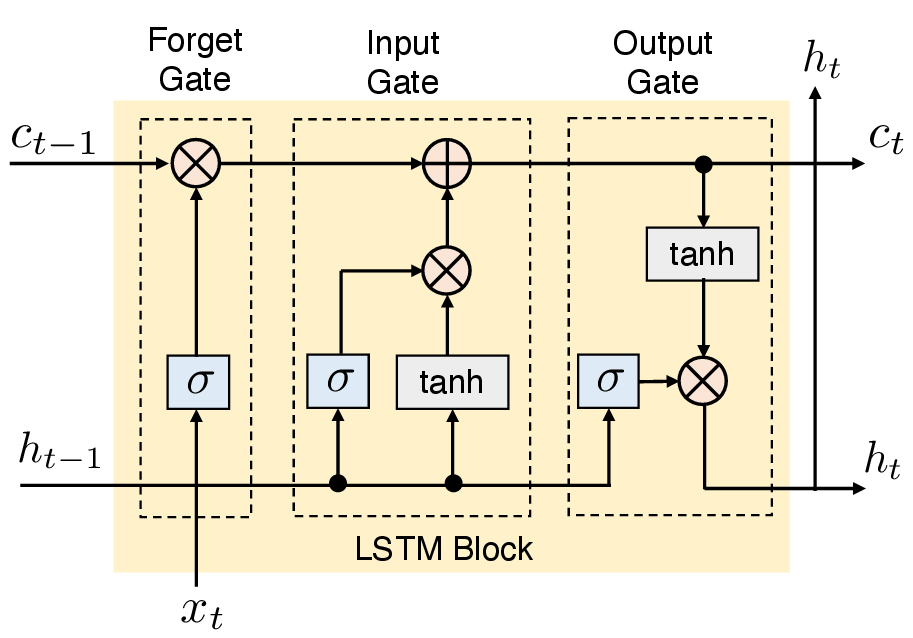}
         \caption{LSTM with a forget gate.}
         \label{fig:lstm}
     \end{subfigure}
     \\ \vspace{3mm}
     \begin{subfigure}[b]{1.0\linewidth}
         \centering
         \includegraphics[width=0.8\linewidth]{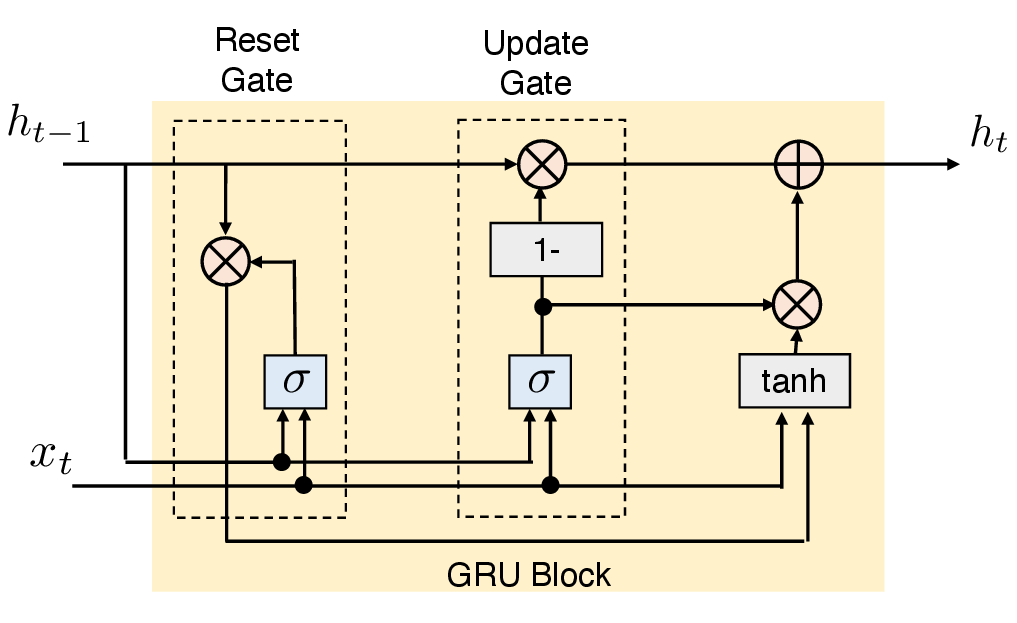}
         \caption{GRU.}
         \label{fig:gru}
     \end{subfigure}
    \caption{Comparison of RNN, LSTM, and GRU architectures \cite{yu2019review}.}
    \label{fig:rnn_comp}
\end{figure}

\Ac{lstm} is one of the most popular RNN models that can reduce the effects of vanishing and exploding gradients \cite{hochreiter1997long} \cite{gers2000learning} by introducing a gating mechanism to input or forget certain features. Another popular model is a \ac{gru} \cite{cho2014learning}, in which recurrent units adaptively capture dependencies of different time scales to accommodate the higher memory requirements of LSTM. A comparison of these architectures is shown in Fig.~\ref{fig:rnn_comp}.
For more details of RNN, in particular LSTM, please refer to \cite{salehinejad2017recent,yu2019review,van2020review}.

The LSTM and GRU models have been widely applied to communication systems. In particular, due to the analogous structure to convolutional codes, RNNs may be well suited for decoding of convolutional codes. Similarly, RNNs have been successfully applied to signal detection for channels with memory \cite{farsad2018neural}.

\subsubsection{Graph Neural Network}
\label{sec:gnn}
While ML effectively captures hidden patterns in Euclidean data, a common assumption of existing ML algorithms is that instances are independent of each other. This assumption no longer holds for graph data, where every node is related to others. Extending DNN models to non-Euclidean domains, which is generally referred to as geometric DL, has been an emerging area of research.
In particular, a \ac{gnn} that operates on the graph domain has recently become a popular graph analysis method \cite{scarselli2008graph,xu2018powerful,zhou2020graph,wu2020comprehensive}. 

Let $G \in (\mathcal{V}, \mathcal{E})$ be a graph, where $\mathcal{V}$ is the node set and $\mathcal{E}$ is the edge set. Let $\mathcal{N}_u$ be the neighborhood of some node $u \in \mathcal{V}$. Additionally, let $\mathbf{x}_u$ be the properties of node $u \in \mathcal{V}$. GNN implements a permutation-equivalent layer, called a GNN layer, which maps a representation of a graph into an updated representation of the same graph. Although the design of GNN layers is one of the active research areas, one popular approach is \ac{mpnn} layers (other popular approaches include graph convolutional networks \cite{kipf2016semi} and graph attention networks \cite{velivckovic2017graph}).
In an MPNN layer in a generic GNN, nodes update their representations by aggregating the messages received from their immediate neighbors \cite{bronstein2021geometric} and the output of the layer (node representations $\mathbf{h}_u$ for each $u \in \mathcal{V}$) is expressed as
\begin{align}
    \mathbf{h}_u = \phi \left(
    \mathbf{x}_u, \bigoplus_{v \in \mathcal{N}_u} \psi (\mathbf{x}_u, \mathbf{x}_v)
    \right),
\end{align}
where $\phi$ and $\psi$ are typically trainable differential functions, whereas $\bigoplus$ is a nonparametric permutation invariant aggregation operator that can take an arbitrary number of inputs. In particular, $\phi$ and $\psi$ are referred to as update and message functions, respectively.

Due to its close relationship to a Tanner graph, a GNN is particularly useful for designing and decoding codes over graph. One of the major advantages of GNN is their scalability, i.e., a GNN trained for a small code length will generalize to any code length, while this usually requires additional training.

\subsubsection{Transformer}
\label{sec:transformer}
\begin{figure}[t]
\centering
\includegraphics[width=0.85\linewidth]{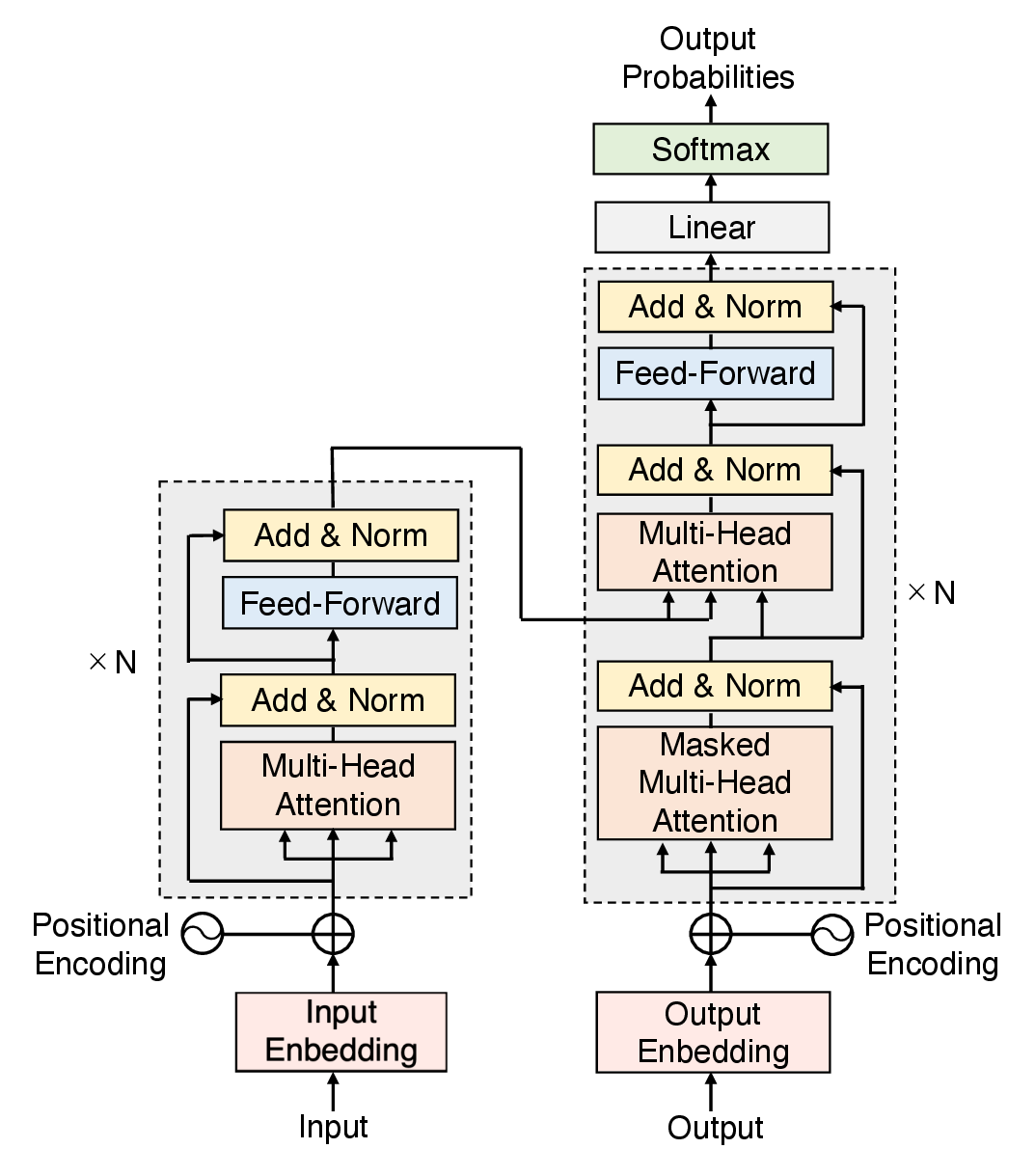}
\caption{The Transformer architecture \cite{vaswani2017attention}.}
\label{fig:transformer} 
\end{figure}

Transformer is a DL architecture based on the multi-head attention mechanism \cite{vaswani2017attention}. As depicted in Fig.~\ref{fig:transformer}, it consists of an encoder and a decoder, each of which has several Transformer blocks having the same architecture. Each Transformer block consists of a multi-head attention layer, a feed-forward neural network, a shortcut connection, and a layer normalization. Given a sequence of elements, the self-attention mechanism explicitly models the dependencies among all entities of a sequence.

It has no recurrent units, and thus requires less training time than previous RNN architectures, such as LSTM, and its later variant has been widely adopted for training most of representative LLMs, such as Open AI's \ac{gpt} series \cite{brown2020language}, Meta's \ac{llama} \cite{touvron2023llama}, Googles \ac{palm} \cite{chowdhery2023palm}, and Gemini \cite{team2023gemini} are based on the Transformer model. More details and the applications of the Transformer can be found in \cite{lin2022survey,khan2022transformers,han2022survey}.

Although Transformer was originally proposed for NLP tasks \cite{vaswani2017attention}, it has been successfully adopted for a variety of tasks such as computer vision \cite{dosovitskiy2020image}, audio applications \cite{radford2023robust}. For the physical layer technologies, the use of Transformer is rather new \cite{wang2022transformer}. Due to its excellent performance, Transformer has the potential to improve the performance of existing DL methods for communication engineering problems.

\subsubsection{Diffusion Model}
\label{sec:diffusion}
\Acp{dm} are a class of probabilistic generative models that progressively corrupt data by injecting noise, and then learn to reverse this process for sample generation. 
The training procedure consists of two phases: the forward diffusion process and the backward denoising process \cite{chang2023design}.
In the forward process, typically Gaussian noise is injected into the training data until it becomes pure Gaussian.
In the backward process, the noise is sequentially removed to reconstruct the original image. The noise subtracted at each step is estimated by a neural network.
Among different formulations \cite{yang2023diffusion,cao2024survey}, \ac{ddpm} \cite{ho2020denoising} is a representative DM inspired by the theory of non-equilibrium thermodynamics.

Due to its high generative quality and versatility, DM could be applied to many problems in communication systems, such as channel estimation \cite{fesl2024diffusion}, signal detection \cite{letafati2023denoising}, AE \cite{letafati2023probabilistic,kim2023learning,wu2024cddm}, and network optimization problems \cite{du2023beyond,bariah2024large}. However, the application of DMs to the physical layer is a relatively new area of research \cite{letafati2023wigenai,van2023generative,zhao2024generative}.

\section{DL for Code Design}
\label{sec:design}

\begin{table*}[h]
    \centering
    \caption{Summary of DL-based polar code design.}
    \begin{tabular}{|p{4cm}|p{3cm}|p{10cm}|} \hline
        Category & Reference & Main Contributions \\ \hline

        \multirow{5}{*}{Advanced Decoding Schemes} &
        Ebada et al. \cite{ebada2019deep} & 
        Design for BP decoding with finite iteration count. \\ \cline{2-3}

        &
        Huang et al. \cite{huang2019ai} & 
        RL-based design of polar codes for SCL decoding. \\ \cline{2-3}

        &
        Leonardon et al. \cite{leonardon2021using} & 
        Design that minimizes BLER under SCL decoding via projected gradient descent. \\ \cline{2-3}
        
        &
        Liao et al. \cite{liao2023scalable} & 
        GNN-based polar code construction for CA-SCL decoding. \\ \cline{2-3}
        
        &
        Miloslavskaya et al. \cite{miloslavskaya2023neural} & 
        Optimization of polar codes with dynamic frozen bits under SCL decoding.
        \\ \hline

        \multirow{3}{*}{Nested Polar Codes} &
        Huang et al.  \cite{huang2019ai} & 
        Construction of nested polar codes via advantage actor-critic  algorithms. \\ \cline{2-3}
        
        &
        Li et al. \cite{li2021learning} & 
        Stochastic policy optimization by a customized network. \\ \cline{2-3}

        &
        Ankireddy et al. \cite{ankireddy2024nested} & 
        Nested polar code construction based on sequence modeling and Transformer. \\ \hline

        Polar Codes with Large Kernel & 
        Hebbar et al. \cite{hebbar2024deeppolar} & 
        Polar codes via large nonlinear neural network-based kernels. \\ \hline

        PAC Codes &
        Mishra et al. \cite{mishra2022modified} & 
        RL-based algorithm for rate-profile construction of PAC codes. \\ \hline

    \end{tabular}
    \label{tab:polar_design}
\end{table*}

Modern capacity-approaching codes such as LDPC and polar codes are usually designed based on well-established analytical tools, such as \ac{de} \cite{richardson2001design,mori2009performance} and \ac{exit} chart \cite{ten2001convergence}. However, these techniques rely on assumptions that do not hold in practice. For example, for the design of LDPC codes, DE and EXIT chart analyses assume simple channel models, such as \ac{biawgn} channels, infinite code length, and unlimited \ac{bp} decoding iterations. These techniques are also used for polar code design, but they are again limited to simple channel models and decoding schemes such as \ac{sc} decoding. For more realistic channel models and advanced decoding schemes, DL could replace or support existing code design techniques. We have summarized the DL-based approaches to polar code design in Table~\ref{tab:polar_design}.

\subsection{LDPC Code Design}
\label{sec:LDPCDesign}
Irregular LDPC codes are characterized by a variable degree distribution $\lambda(x)$ and a check degree distribution $\rho(x)$, which are expressed as
\begin{align}
    \lambda(x) = \sum^{d_\text{v}}_{i=2} \lambda_i x^{i-1}, \ 
    \rho(x) = \sum^{d_\text{c}}_{i=2} \rho_i x^{i-1},
\end{align}
where $\lambda_i$ and $\rho_i$ represent the fraction of edges emanating from \acp{vn} and \acp{cn} of degree $i$ and $\lambda(1)=\rho(1)=1$.
The maximum variable degree and check degree are denoted by $d_\text{v}$ and $d_\text{c}$, respectively. 
The degree distribution is often optimized to maximize the iterative decoding threshold, which is defined as the lowest channel \ac{snr} at which the message distribution in BP evolves in such a way that its associated error probability converges to zero as the number of iterations tends to infinity. The method of identifying a threshold by tracking the evolution of the message distribution is termed DE \cite{richardson2001design}.

The code design problem belongs to the class of nonlinear constraint satisfaction problems with continuous space parameters, where we first explore the space of degree distributions to find degree distribution pairs, traditionally solved by differential evolution \cite{richardson2001design}, and then evaluate the BP threshold of the selected pairs via DE.
In \cite{nisioti2020design}, the authors modeled the code design process as a supervised learning problem by mapping the recursive update equation of DE to an RNN architecture, which they refer to as neural density evolution (NDE). They also proposed a multi-objective loss function for NDE that ensures its high configurability, i.e., various code rates and maximum degrees. Their simulations show that the proposed designs achieve the performance of state-of-the-art designs in asymptotic settings for a variety of codeword lengths and channels.

\subsection{Polar Code Design}
\label{sec:PolarDesign}
An encoder of polar codes of length $N$ is represented by the generator matrix $\mathbf{G}_{N} = \left(\begin{smallmatrix}1&0\\1&1\end{smallmatrix}\right)^{\otimes n}
\in \mathbb{F}_{2}^{N \times N}$, where $n=\log_2{N}$ is a positive integer and $\mathbf{A}^{\otimes n}=\mathbf{A} \otimes \mathbf{A}^{\otimes(n-1)}$ is the $n$th Kronecker power of the matrix $\mathbf{A}$ \cite{arikan2009channel}. Let $\mathcal{I} \subset \{0, 1, \ldots, N-1\}$ denote a set of information bit indices with its cardinality $K = |\mathcal{I}|$, and  $\mathcal{F} = \{0, 1, \ldots, N-1\} \backslash \mathcal{I}$ denote the complement of $\mathcal{I}$ with its cardinality $|\mathcal{F}|=N-K$. Letting $\mathbf{u} = (u_0, u_1, \ldots, u_{N-1}) \in \mathbb{F}_{2}^{N}$ be an input vector to the polar encoder, the bits $u_i$ with $i\in \mathcal{I}$ are chosen to carry information, whereas those with $i \in \mathcal{F}$ are frozen (i.e., fixed to a predetermined bit value known by encoder and decoder). The code rate of the polar code is thus $R = K/N$. Polar code design or construction is equivalent to identifying an appropriate index set $\mathcal{I}$ for a given channel model and decoding scheme.

In his original paper, Ar{\i}kan suggested using Monte-Carlo simulations to estimate the reliabilities of bit channels \cite{arikan2009channel}. Subsequently, DE \cite{mori2009performance} and its improved version \cite{tal2013construct} were proposed to accurately estimate the reliabilities at the cost of high complexity. This complexity was alleviated by \ac{ga} of DE \cite{trifonov2012efficient}, improved GA \cite{ochiai2020capacity}, and \ac{rca} \cite{ochiai2022new}.

\subsubsection{Design for Advanced Decoding Schemes}
Although polar codes were originally proposed with SC decoding \cite{arikan2009channel}, their finite-length performance is unsatisfactory. In order for polar codes to achieve performance comparable to other capacity-approaching codes, advanced decoding schemes, such as \ac{scl} or \ac{cascl} decoding \cite{tal2015list}, are required. However, the above-mentioned polar code construction schemes assume SC decoding and there is no explicit approach to designing polar codes for SCL and CA-SCL decoding. 

In \cite{ebada2019deep}, the authors proposed to design polar codes for BP decoding with limited number of iterations over both \ac{awgn} and Rayleigh fading channels. By representing the frozen and non-frozen bit vectors by soft-valued vectors, which can be considered as training weights of a neural network, training is performed to minimize the cross entropy loss between transmitted and estimated codewords while satisfying the target code rate requirement and training convergence. The simulations showed that the learned polar code outperforms the performance of the 5G polar code under Arıkan's conventional BP decoder.

On the other hand, the authors in \cite{huang2019reinforcement,huang2019ai} proposed a genetic algorithm and RL-based design of polar codes for SCL decoding. As a reward function in RL, the authors computed an SNR required for a code to achieve a target \ac{bler} via Monte-Carlo simulations. In the same line of research, the authors in \cite{liao2021construction} proposed a tabular RL-based construction of polar codes for SCL decoding. Instead of evaluating the BLER based on the Monte-Carlo method as in \cite{huang2019reinforcement,huang2019ai}, they designed the reward function that sends a negative immediate reward (penalty) to the agent when the selected action causes a frame error in genie-aided SCL decoding\footnote{In genie-aided SCL decoding, the decoder can always output the correct codeword if it is in the list.}. The proposed method achieved comparable or slightly better performance with a lower computational complexity in training than the method in \cite{huang2019ai}.

As another approach, the authors in \cite{leonardon2021using} proposed a two-step optimization method. More specifically, they first trained MLPs to predict BLER under SCL decoding from input frozen bit sequences, and then the code that minimizes the BLER was constructed via \ac{pgd} \cite{madry2017towards} which has been widely studied in the realm of adversarial attacks on neural networks. Simulations demonstrated that the proposed construction successfully improves the performance of the codes on the dataset used for predicting BLER.

More recently, the authors in \cite{liao2023scalable} proposed a GNN-based polar code construction algorithm for CA-SCL decoding. More specifically, a polar code is first mapped onto a unique heterogeneous graph called the \ac{pccmp} graph, and then a heterogeneous GNN-based iterative message-passing algorithm is proposed which aims to find a PCCMP graph corresponding to the polar code with minimum BLER under CA-SCL decoding. The proposed GNN-based iterative message-passing method has a salient property that a single trained model can be directly applied to constructions for different design SNRs and different block lengths without any additional training. Numerical experiments showed that the proposed constructions outperform classical constructions in \cite{tal2013construct} under CA-SCL decoding.

In \cite{miloslavskaya2023neural}, the authors proposed neural network-based adaptive polar coding scheme that adapts to various channel conditions and quality of service requirements. Specifically, the authors developed an MLP-based performance prediction framework for polar codes with dynamic frozen bits under SCL decoding. Then the authors presented a new class of polar codes with dynamic frozen bits parameterized by a single integer parameter, and used the performance prediction framework to optimize the parameter for a given target BLER, list size, code length and rate. The simulation results show that the proposed codes outperform 5G polar codes under CA-SCL decoding with various list sizes.
Although a neural network is not used due to the training difficulty, the same authors also proposed an RL-based method to design dynamic frozen bits of polar codes that minimize the BLER under SCL decoding in \cite{miloslavskaya2023design}.

\subsubsection{Nested Polar Code Construction}
In general, on-the-fly design of polar codes that adaptively select frozen bits for a given channel may be too complex to implement in practical systems. On the other hand, the authors in \cite{schurch2016partial,bardet2016algebraic,he2017beta,mondelli2018construction} studied the design of polar codes based on a universal reliability order of bit channels that is independent of channel conditions. As such, it is preferable in practice to impose a nested property on polar codes with different rates, so that all polar codes can be derived from the same mother code based on the universal reliability sequence \cite{bioglio2020design}.

\begin{figure}[t]
\centering
\includegraphics[width=0.95\linewidth]{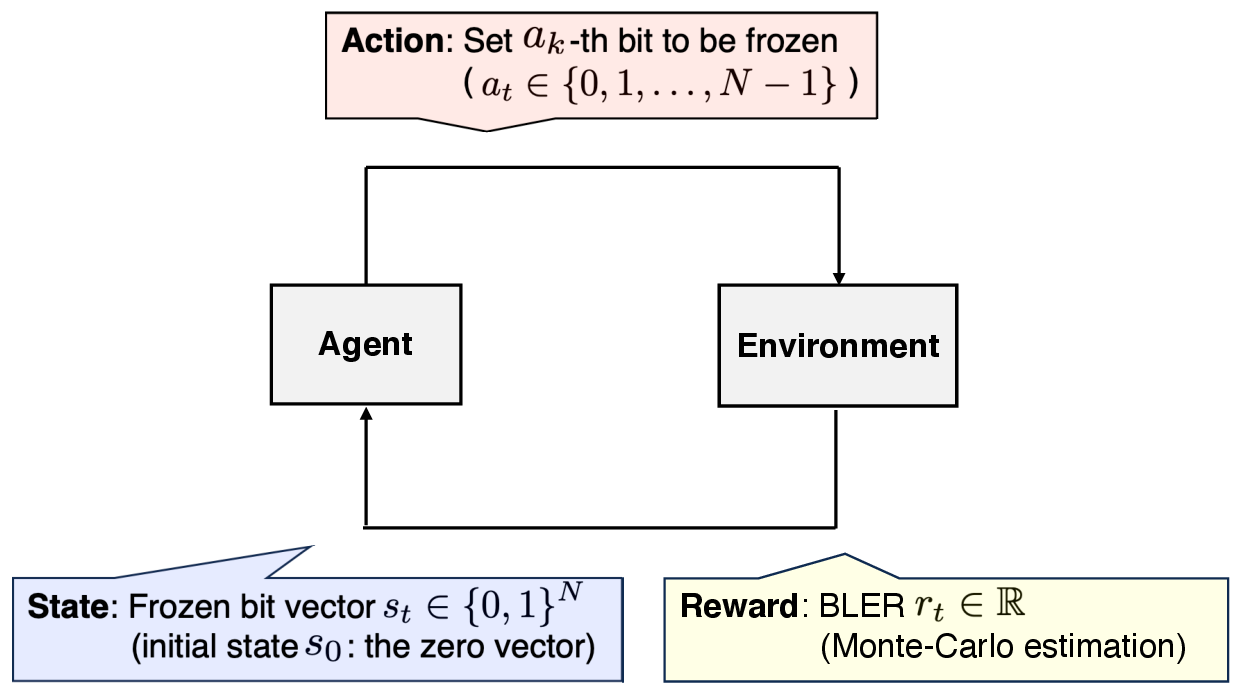}
\caption{RL-based construction of a rate-$K/N$ nested polar code at time $t \in \{1, \ldots, K\}$ in \cite{huang2019reinforcement,huang2019ai}. The goal is to minimize the expected cumulative BLER, i.e., $\sum^{K}_{t=1} r_k$. The ones in the frozen bit vector $s_t$ indicate locations of frozen bits.}
\label{fig:rl_polar} 
\end{figure}

The authors in \cite{huang2019reinforcement,huang2019ai} proposed constructing nested polar codes via \ac{a2c} algorithms \cite{konda1999actor}.
The paper regarded code construction as a multi-step MDP, where for a given $(N, K)$ polar code (current state), a new action is taken to construct $(N, K + 1)$ polar code (an updated state). This MDP is illustrated in Fig.~\ref{fig:rl_polar}. The reliability ordered sequence is then constructed by sequentially appending the actions to the end of initial polar code construction. The proposed code design was shown to outperform the conventional DE construction under SCL decoding.

Meanwhile, the authors in \cite{li2021learning} first transformed the problem of nested polar code construction into a stochastic policy optimization problem for sequential decision, and then represented the policy by a customized neural network. Furthermore, the authors proposed a gradient-based algorithm to minimize the average loss of the policy. Simulation results demonstrate that the proposed construction achieves better performance than the state-of-the-art nested polar codes for SCL decoding in \cite{huang2019reinforcement,huang2019ai}.
A similar construction of nested polar codes has also been proposed in \cite{ankireddy2024nested}, where the authors parameterized the
policy network by a Transformer encoder-only model, which can directly predict the next information bit in the nested sequence. It was shown that the proposed Transformer-based construction can achieve better error rate performance than the approach proposed in \cite{li2021learning}.

\subsubsection{Design of Polar Codes with Large Kernel}
Another way to improve finite-length performance of binary polar codes is to increase the size of the polarization kernel. 
In fact, channel polarization holds for all kernels provided that they are not unitary and not upper triangular under any column permutation \cite{korada2010polar}. However, binary polar codes with large kernels exhibit poor performance for practical short-to-medium block lengths and face an exponential increase in computational complexity with kernel size.

In \cite{hebbar2024deeppolar}, the authors proposed polar codes via large nonlinear neural network-based kernels, termed as DEEPPOLAR, and its decoder based on a generalization of SC decoding. They also developed a principled curriculum-based training methodology that allows DEEPPOLAR to generalize well to high SNR scenarios, characterized by rare error events. It was shown that the DEEPPOLAR outperforms the classical polar codes with SC decoding.

\subsubsection{Design of Polarization Adjusted Convolutional Codes}
In the Shannon Lecture at the 2019 \ac{isit}, Ar{\i}kan introduced a new class of codes, called \ac{pac} codes, that concatenate  convolutional precoding with the polar transform \cite{arikan2019sequential}. PAC codes significantly improve the performance of polar codes at short-to-moderate block lengths, where channel polarization occurs relatively slowly.

The first step in encoding of PAC codes is rate profiling. For PAC codes of rate-$K/N$, this step inserts $K$ information bits into a vector of length $N$, which is subsequently input to the convolutional precoder. The selection of $K$ bit indices out of $N$ possible indices is called rate profile construction and its design significantly affects the performance of PAC codes.

In \cite{mishra2022modified}, the authors proposed an RL-based algorithm for rate-profile construction of PAC codes.
Specifically, by mapping the rate profile construction problem to MDP, the authors proposed Q-learning with a set of customized reward and update strategies. Simulation results showed that the proposed rate-profile construction provides better error rate performance compared to the Monte-Carlo-based rate profiling design in \cite{moradi2021monte}.

\section{DL for Channel Decoding}
\label{sec:decoding}
DL methods for channel decoding have been an active area of research and have been extensively studied as a means to replace or assist conventional decoding algorithms. In what follows, we first review model-free decoders that do not assume specific code structure and thus are applicable to any codes. We then review model-based DL BP decoding methods take take into account specific factor graphs. Subsequently, we focus on DL methods for decoding polar codes, convolutional/turbo codes, and cyclic codes.

\subsection{Model-Free Decoders}
\label{sec:model-free}

\begin{table*}[h]
    \centering
    \caption{Summary of model-free decoders.}
    \begin{tabular}{|p{4cm}|p{3.6cm}|p{9cm}|} \hline
        Category & Reference & Main Contributions \\ \hline

        \multirow{4}{*}{MLP Decoders} &
        Gruber et al. \cite{gruber2017deep} & 
        Initial work on MLP-based channel decoding. \\ \cline{2-3}

        &
        Seo et al. \cite{seo2018decoding} & 
        Investigation on the impact of various configurations of an MLP decoder. \\ \cline{2-3}

        &
        Leung et al. \cite{leung2019low} & 
        Empirical study on the impact of hyperparameters of the MLP decoder. \\ \cline{2-3}
               
        &
        Leung et al. \cite{leung2022low} & 
        Investigated small MLP for applications with low energy and latency. \\ \hline
        
        \multirow{5}{*}{Advanced DL Models} &
        Lyu et al. \cite{lyu2018performance} & 
        Investigation on different types of DL decoders, namely, MLP, CNN, RNN. \\ \cline{2-3}
                       
        &
        Sattiraju et al. \cite{sattiraju2018performance} & 
        Bi-GRU-based decoder. \\ \cline{2-3}
        
        &
        Zhu et al. \cite{zhu2020learning} & 
        Residual MLP, CNN, RNN-based decoders. \\ \cline{2-3}
        
        &
        Choukroun et al. \cite{choukroun2022error} & 
        Novel Transformer architecture for decoding block codes. \\ \cline{2-3}

        &
        Choukroun et al. \cite{choukroun2022denoising} & 
        DDPM for soft-decision decoding of linear codes. \\ \hline
        
        \multirow{3}{*}{Syndrome-Based Loss Function} & 
        Bennatan et al. \cite{bennatan2018deep} & 
        Syndrome-based approach to soft-decision decoding of linear codes. \\ \cline{2-3}

        &
        Kamassury et al. \cite{kamassury2020iterative} & 
        Iterative algorithm, referred to as iterative error decimation. \\ \cline{2-3}
 
        &
        Artemasov et al. \cite{artemasov2023soft} & 
        SISO decoder based on Stacked-GRU for turbo product codes. \\ \hline
       
        \multirow{4}{*}{Adaptability} &
        Wang et al. \cite{wang2018unified} & 
        Unified DL-based decoder for polar and LDPC codes. \\ \cline{2-3}

        &
        Jiang et al. \cite{jiang2019mind} & 
        A meta learning-based model independent neural decoder. \\ \cline{2-3}
       
        &
        Lee et al. \cite{lee2020training} & 
        Transfer learning for decoding a set of rate-compatible polar codes. \\ \cline{2-3}
               
        &
        Artemasov et al. \cite{artemasov2023unified} & 
        A unified DL decoder for BCH and polar codes concatenated with CRC. \\ \hline
        
        \multirow{2}{*}{RL-Based Approach} & 
        Carpi et al. \cite{carpi2019reinforcement} & 
        DQN for iterative bit-flipping decoding of binary linear codes. \\ \cline{2-3}

        &
        Gao et al. \cite{gao2021reinforcement} & 
        Q-learning-based bit-flipping decoding for polar codes. \\ \hline

        \multirow{2}{*}{Complexity Reduction} &
        Kavvousanos et al. \cite{kavvousanos2018simplified,kavvousanos2019hardware,kavvousanos2021optimizing,kavvousanos2022low} & 
        Magnitude-based pruning and quantization for parameter reduction. \\ \cline{2-3}

        &
        Cavarec et al. \cite{cavarec2020learning} & 
        A DL-aided adaptation of the order parameter in OSD. \\ \hline

        \multirow{4}{*}{Other Approaches} &
        Raviv et al. \cite{raviv2020perm2vec} & 
        Data-driven framework for permutation selection in permutation decoding. \\ \cline{2-3}

        &
        Kurmukova et al. \cite{kurmukova2024friendly} & 
        Friendly jamming for improving decoding performance.  \\ \cline{2-3}
        
        &
        Tsvieli et al. \cite{tsvieli2023learning} & 
        Investigation on the problem of maximizing the margin of the decoder. \\ \cline{2-3}
        
        &
        Zhong et al. \cite{zhong2020deep,zhong2023deep} & 
        DL-based decoders for spin-torque transfer magnetic random access memory. \\ \hline
        
    \end{tabular}
    \label{tab:model-free}
\end{table*}

A model-free decoder employs neural networks that do not take into account any specific structure of the codes and thus can potentially benefit from the powerful architectures of advanced DL models. However, such decoders typically suffer from the curse of dimensionality, since the size of the training dataset grows exponentially with the number of information bits. 
On the other hand, a potential advantage over conventional non-DL-based decoding is a highly parallelizable structure, allowing \emph{one-shot} decoding instead of iterative decoding. The model-free approaches that we will discuss below is summarized in Table~\ref{tab:model-free}.

\subsubsection{MLP Decoders}
The paper \cite{gruber2017deep} is one of the initial works on DL-based channel decoding where the authors investigated the direct application of MLP to decoding of random and polar codes.
Their empirical results demonstrated that for structured codes, the DL decoder can generalize even for codewords not seen in the training phase, and the DL decoder can achieve \ac{map} decoding performance for a very small code lengths such as $16$ bits, but learning for longer codes is prohibitively complex due to the exponentially increasing training complexity.

Meanwhile, the authors in \cite{seo2018decoding} investigated the impact of various configurations of an MLP decoder on BER performance, such as the number of hidden layers, the number of nodes for each layer, and activation functions. Similarly, the paper \cite{leung2019low} empirically studied the impact of the number of hidden layers and nodes as well as a training SNR of the MLP decoder on its performance and investigated the minimum numbers required to achieve similar performance to the optimal maximum-likelihood decoder for short linear and nonlinear block codes.

In \cite{leung2020multi,leung2022low}, the authors investigated the application of small MLP decoders to low-energy and low-latency applications. In particular, the paper made comparisons between single-label and multi-label neural decoders, and demonstrated that the multi-label decoder generally requires more hidden layers and nodes to achieve similar performance to the single-label decoder.

\subsubsection{Advanced DL Models}
The performance of the above-mentioned MLP decoder can be enhanced by advanced DL models. For instance, in \cite{lyu2018performance}, the authors investigated different types of decoders based on MLP, CNN, and RNN (in particular, LSTM). Their empirical results demonstrated that the RNN and CNN decoders can actually achieve better BER performance than the MLP decoder at the cost of higher computation time. In \cite{sattiraju2018performance}, RNN (\ac{bigru}) was used for encoding/decoding of turbo codes as in \cite{kim2018communication}.

To further improve the performances, the authors of \cite{cao2020learning,zhu2020learning} introduced the concept of residual learning \cite{he2016deep} to the MLP, CNN, and RNN-based decoders. Specifically, the paper introduced a denoiser network prior to the decoder that simply aims to remove noise induced at the channel, and proposed a training loss that considers both denoising and decoding performance. It was demonstrated that the proposed denoiser network can improve the BER performance at the cost of marginal increase in run time.

More recently, inspired by the success of the Transformer model in various applications \cite{vaswani2017attention}, a novel Transformer architecture for decoding algebraic block codes, termed \ac{ecct} was proposed in \cite{choukroun2022error}. 
The ECCT takes as input a concatenation of reliabilities of codeword bits (absolute values of received symbols in the case of BI-AWGN) and syndrome bits as its input where each element of which is embedded in a high-dimensional space with its own position-dependent embedding vector.
Then, a self-attention mechanism is employed, where the interaction between bits specified by the code structure, i.e., \ac{pcm}, is incorporated as domain knowledge. 
Extensive simulation results demonstrated that the proposed Transformer-based decoder outperforms state-of-the-art neural decoders.

Although the ECCT in \cite{choukroun2022error} employs a mask matrix that is derived from the PCM, there exist numerous PCMs for the same code which will lead to different decoding performances. Motivated by this fact, the authors in \cite{park2023mask} addressed the problem of identifying the optimal PCM. In particular, the authors proposed a systematic mask matrix constructed from the systematic PCM which results in sparse self-attention map, and proposed a novel Transformer architecture called a double-masked ECCT that consists of two parallel masked self-attention blocks employing distinct mask matrices. 

Meanwhile, in \cite{choukroun2022denoising}, the authors employed DDPM \cite{ho2020denoising} for soft decoding of linear codes with arbitrary block lengths. Their framework models the transmission over the AWGN channel as a series of diffusion steps that can be iteratively reversed. The paper also proposed to condition the diffusion decoder on the number of parity check errors and to employ a line-search procedure to control the reverse diffusion step size.

Furthermore, in \cite{choukroun2024foundation}, the authors proposed a foundation model \cite{bommasani2021opportunities} for channel codes by extending the Transformer architecture. A foundation model refers to a model that is initially trained on a wide range of data, generally based on self-supervision, and then adapted (e.g., transferred or fine-tuned) to a wide range of downstream tasks. Thus, the proposed framework provided a universal decoder that is capable of adapting and generalizing to any (unseen) code of any length.

\subsubsection{Syndrome-Based Loss Function}
\label{sec:synd}
Syndrome decoding is a well-known approach for decoding algebraic codes. Several approaches have been proposed for training DL-based syndrome decoders that estimate the transmitted codeword from the syndrome. Syndrome-based training does not rely on the knowledge of the transmitted codeword, and is thus promising for online adaptation to changing channel conditions.

\begin{figure}[t]
\centering
\includegraphics[width=0.99\linewidth]{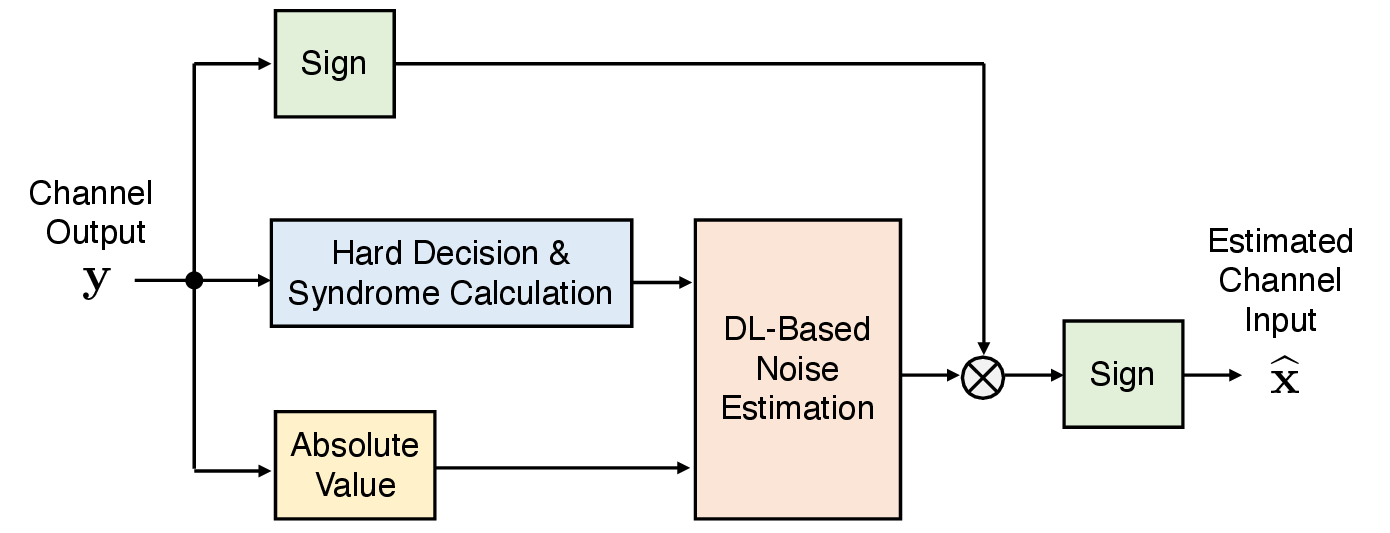}
\caption{The syndrome-based DL decoder proposed in \cite{bennatan2018deep}. The channel input sequence $\mathbf{x}$ consists of \ac{bpsk} symbols, and $\mathbf{y}$ is the output of a \ac{biso} channel.}
\label{fig:syndrome} 
\end{figure}

The paper \cite{bennatan2018deep} is one of the early works on syndrome-based DL decoding, which proposed to use the absolute values of the received symbols, i.e., reliabilities in the case of the BI-AWGN channel, and the syndrome of its hard decisions for decoding, instead of directly using the received symbols as the input to the decoder. The proposed decoder is illustrated in Fig.~\ref{fig:syndrome}. Furthermore, the authors introduced permutations from the code's automorphism group \cite{macwilliams1977theory,dimnik2009improved} as a preprocessing. Permutations in this group have the property that the permuted version of any codeword is guaranteed to be also a valid codeword, i.e., the permuted input of the decoder is a noisy valid codeword. Simulations demonstrated that the proposed framework can achieve near maximum-likelihood performance for short \ac{bch} codes.

The syndrome-based approach, which attempts to predict the error vector from its syndrome alone, can suffer from the potential presence of inconsistent training examples, called \emph{disturbance} \cite{tallini1995neural}, i.e., training examples with the same syndromes but with different error vectors. To solve this problem, the authors in \cite{kamassury2020iterative} proposed an iterative algorithm, referred to as iterative error decimation, which is robust against the superposition of error patterns. In each iteration, the DL decoder estimates the error vector and then decimates (subtracts) it from the received vector. The simulation results demonstrated that the proposed scheme improves the performance of the scheme in \cite{bennatan2018deep}.

Furthermore, the authors in \cite{artemasov2023soft} proposed a syndrome-based approach to \ac{siso} decoding of BCH component codes in turbo product codes \cite{pyndiah1998near} based on Stacked-GRU, which is an RNN architecture composed of GRU. They introduced a regularization term into a loss function and demonstrated that the proposed DL decoder outperforms the original chase decoder in \cite{pyndiah1998near}.

\subsubsection{Adaptability}
Adaptive coding is a technique for adapting a code rate to a channel condition in wireless channels. For a DL-based decoder, supporting multiple code rates not only requires multiple training, which is computationally intensive, but also requires a large amount of memory to store the learned DL parameters.

To address this issue, a unified DL-based decoder for polar and LDPC codes was proposed in \cite{wang2018unified}, which supports different codes, i.e., polar and LDPC codes, with a single DNN by utilizing a code indicator at the decoder input. The simulation results demonstrated the potential of the unified decoder for very short code lengths, e.g., $16$ bits.
A similar unified DL decoder using the code indicator was also proposed in \cite{artemasov2023unified} for BCH and CRC-concatenated polar codes. Their results demonstrated that, for a code length of $64$, the proposed unified decoding scheme with code indicator achieves a small performance gap of less than $0.2$ dB from the decoder trained solely for the single code. 

Meanwhile, the paper \cite{jiang2019mind} introduced the meta-learning-based neural decoder, termed as \ac{mind}, which can adapt to a new channel with a small number of pilots and few gradient descent steps. Specifically, the proposed approach consists of meta-learning and meta-training steps, where the model learns good initial parameters in the meta-learning step, and then adapts the parameters to the observed channel in the meta-testing phase using minimal adaptation data and pilots. It has been demonstrated that the proposed scheme can adapt to a channel while achieving a performance close to that of a DL decoder designed solely for the particular channel. In the same line of research, the authors in \cite{lee2020training} proposed transfer learning to efficiently train decoders for a set of rate-compatible polar codes that are expurgated from the same mother code as in 5G NR.

\subsubsection{RL-Based Approaches}
Although the majority of previous works on DL-based channel decoding are based on supervised learning, several RL-based approaches have been proposed for the cases where supervisory data (ground truth) is unavailable.

The paper \cite{carpi2019reinforcement} is one of the earliest works on RL-based channel decoding. Unlike these studies, the authors in \cite{carpi2019reinforcement} proposed an RL framework for iterative \ac{bf} decoding of binary linear codes. Specifically, they linked the BF decision at each step to MDP and applied RL to find good decision strategies. The authors also exploited the permutation automorphism group to improve the performance. The extensive simulations showed that the learned BF decoders with DQN can achieve near-optimal performance for short, high-rate codes. 

Later, the authors in \cite{gao2021reinforcement} applied RL-based BF decoding to polar codes. In contrast to the DQN proposed in \cite{carpi2019reinforcement}, they used simple Q-learning and attempted to map channel observations directly onto estimated codewords. Simulation results showed that the proposed Q-learning achieves comparable performance to the learned BF decoding in \cite{carpi2019reinforcement} with lower complexity.

\subsubsection{Complexity Reduction}
In general, decoding longer codes requires a larger neural network size. Such a network not only requires a huge amount of computational resources in the training phase, but also imposes high computational and space complexities in the inference phase. In particular, the complexity in the inference phase is of practical importance since the training is usually performed offline.

The authors in \cite{kavvousanos2018simplified} attempted to reduce the parameter size of a DL decoder, by introducing various simplified neural network structures with fewer parameters. 
In \cite{kavvousanos2019hardware}, the same authors further extended their work by introducing the magnitude-based pruning \cite{zhu2017prune} and quantization of the parameters.  The proposed decoder was demonstrated to achieve similar performance to the original system in \cite{bennatan2018deep} even with $80$\% reduction of the network parameters and $8$-bit fixed-point representation.
Furthermore, FPGA implementation of the proposed decoder has been presented in \cite{kavvousanos2021optimizing,kavvousanos2022low}.

Meanwhile, the authors in \cite{cavarec2020learning} considered DL-aided complexity reduction of \ac{osd} \cite{fossorier1995soft}, which is a soft-decision decoding algorithm of linear block codes that approaches the optimal maximum-likelihood decoding performance especially for short codes. Although increasing the order parameter of OSD leads to near-maximum-likelihood performance, it may waste computational resources when the received signal can be decoded with lower order. The paper \cite{cavarec2020learning} proposed a learning-based approach to adapt the required order parameter to the channel condition and demonstrated the effectiveness of the proposed scheme in terms of the performance-complexity trade-off through numerical simulations.

\subsubsection{Other Approaches}
It is known that multiple decoding attempts over different permutations of received codewords provide a performance gain \cite{dimnik2009improved,elkelesh2018belieflist}. However, it remains unclear how to choose the permutation that yields the best performance. To address this, the authors in \cite{raviv2020perm2vec} presented a DL approach to selecting candidates from the code's automorphism group in permutation decoding. In this scheme, a trained network predicts the probability of successful decoding for each permutation, and decoding is performed only for permuted codewords with the highest probabilities. The proposed algorithm has been demonstrated by simulations to achieve remarkable performance gains over a random selection of permutations from the automorphism group.

In \cite{kurmukova2024friendly}, the authors proposed a novel approach referred to as \emph{friendly attack} for improving channel decoding performance, inspired by the concept of adversarial attacks. The proposed scheme introduces small perturbations into the modulated symbols before transmission. The perturbations are designed by a modified iterative fast gradient method \cite{kurakin2018adversarial} such that a loss function between the decoded codeword and the transmitted codeword is minimized. The performance improvement by the proposed scheme has been demonstrated for various codes and decoders. 

Although the use of DL for channel decoding has been experimentally validated, the theoretical justification for the developed algorithm in terms of, e.g. the generalization properties, remains challenging. The authors in \cite{tsvieli2023learning} addressed the problem of maximizing the margin of the decoder for an additive noise channel whose noise distribution is unknown, as well as for a nonlinear channel with AWGN. They formulated a maximum margin optimization problem, which is common in \acp{svm}, for the decoder learning problem, and they relaxed it to a \ac{rlm} problem by several approximation steps. The paper then provided expected generalization error bounds for both models, under optimal choice of the regularization parameter. The paper also presented a theoretical guidance for choosing the training SNR based on the bound for the additive noise channel.

\subsection{DL-Aided BP Decoding}
\label{sec:BP}

\begin{table*}[h]
    \centering
    \caption{Summary of DL-aided BP decoding.}
    \begin{tabular}{|p{4cm}|p{3.2cm}|p{9cm}|} 
        \hline
        Category & Reference & Main Contribution \\ 
        \hline

        \multirow{7}{*}{\shortstack{Neural MS Decoding \\and Its Variants}} &
        Nachmani et al. \cite{nachmani2016learning,nachmani2018deep} & 
        DL-aided BP, NMS, and OMS decoding. \\ \cline{2-3}

        & 
        Lugosch et al. \cite{lugosch2017neural} & 
        NOMS decoding. \\ \cline{2-3}

        &
        Dai et al. \cite{dai2018new} & 
        Neural network-aided OMS and NOMS decoding. \\ \cline{2-3}

        &
        Yu et al. \cite{yu2023neural} & 
        Neural AMS decoding. \\ \cline{2-3}

        &
        Hsu et al. \cite{hsu2024gc} & 
        Neural network-aided VWMS decoding. \\ \cline{2-3}

        &
        Wu et al. \cite{wu2018decoding} & 
        Neural MS decoding with linear approximation for PB-LDPC codes. \\ \cline{2-3}
        
        &
        Kim et al. \cite{kim2024neural} & 
        Neural SCMS decoding. \\ \hline
        
        \multirow{2}{*}{Performance Enhancement} &
        Teng et al. \cite{teng2019convolutional,teng2020convolutional,teng2021convolutional} & 
        CNN-based learned BF for BP. \\ \cline{2-3}

        & 
        Sun et al. \cite{sun2020lstm} & 
        LSTM-based learned BF for BP. \\ \hline

        \multirow{2}{*}{\shortstack{Variants of Random Redundant \\Decoding (RRD)}} &
        Nachmani et al. \cite{nachmani2018deep} & 
        mRRD decoding with RNN-based BP decoders. \\ \cline{2-3}

        & 
        Liu et al. \cite{liu2020deep} & 
        Node-classified redundant decoding algorithm. \\ \hline

        \multirow{3}{*}{Optimization-Based Decoding} &
        Wei et al. \cite{wei2020admm} & 
        Trainable ADMM-penalized decoder. \\ \cline{2-3}

        & 
        Wadayama et al. \cite{wadayama2019deep} & 
        Trainable PGD decoder for LDPC codes. \\ \cline{2-3}

        & 
        Wadayama et al. \cite{wadayama2023proximal} & 
        Proximal decoding for LDPC codes. \\ \hline

        \multirow{2}{*}{RL-Based Approach} &
        Doan et al. \cite{doan2020decoding} & 
        RL-based selection of permutations on factor-graph. \\ \cline{2-3}

        & 
        Habib et al. \cite{habib2020learned,habib2021belief,habib2023reldec} & 
        RL-based scheduling optimization for sequential BP decoding. \\ \hline

        \multirow{2}{*}{Customized Loss Function} &
        Lugosch et al. \cite{lugosch2018learning} & 
        Soft syndrome as a loss function for training a neural BP decoder. \\ \cline{2-3}

        & 
        Teng et al. \cite{teng2020syndrome} & 
        New syndrome losses for syndrome-based DL decoding of polar codes. \\ \cline{2-3}

        & 
        Nachmani et al. \cite{nachmani2022neural} & 
        New loss function based on sparse node and knowledge distillation losses. \\ \hline

        \multirow{11}{*}{Memory/Complexity Reduction} &
        Teng et al. \cite{teng2019low} & 
        Weight quantization mechanism for an RNN polar decoder. \\ \cline{2-3}

        & 
        Ibrahim et al. \cite{ibrahim2022enhanced} & 
        Quantization of an RNN polar decoder. \\ \cline{2-3}
                
        & 
        Xiao et al. \cite{xiao2019finite,xiao2020designing} & 
        Finite alphabet iterative decoders for LDPC codes via quantized RNN. \\ \cline{2-3}
        
        & 
        Lyu et al. \cite{lyu2023optimized} & 
        A joint optimization of message quantization and quantization thresholds. \\ \cline{2-3}
                
        & 
        Lian et al. \cite{lian2018can,lian2019learned} & 
        Weight-sharing across edges based on scalar parameters. \\ \cline{2-3}
                
        & 
        Wang et al. \cite{wang2020model,wang2022normalized} & 
        A parameter sharing scheme within the same layer for a neural NMS decoder. \\ \cline{2-3}
                
        & 
        Dai et al. \cite{dai2021learning} & 
        A weight-sharing scheme for a neural MS decoder of protograph LDPC codes. \\ \cline{2-3}
                
        & 
        Liang et al. \cite{liang2022low,liang2023joint} & 
        Tensor-train and tensor-ring decompositions for parameter size reduction. \\ \cline{2-3}

        & 
        Cheng et al. \cite{cheng2023rate} & 
        A weight-sharing scheme for adapting to multiple code rates. \\ \cline{2-3}

        & 
        Buchberger et al. \cite{buchberger2020pruning} & 
        A novel pruning-based neural BP decoder for short linear block codes. \\ \cline{2-3}

        & 
        Buchberger et al. \cite{buchberger2021learned} & 
        A neural BP with decimation. \\ \hline

        \multirow{3}{*}{GNN Decoders} &
        Satorras et al. \cite{satorras2021neural} & 
        A hybrid inference model that combines BP and GNN. \\ \cline{2-3}

        & 
        Cammerer et al. \cite{cammerer2022graph}  & 
        A fully GNN-based decoder. \\ \cline{2-3}

        & 
        Tian et al. \cite{tian2023scalable} & 
        An edge-weighted GNN decoder. \\ \hline
        
        \multirow{2}{*}{Understanding Neural BP Decoders} &
        Ankireddy et al. \cite{ankireddy2023interpreting} & 
        Empirical study on how the learned weights attenuate the effect of these cycles. \\ \cline{2-3}

        & 
        Adiga et al. \cite{adiga2024generalization} & 
        Theoretical study on the generalization capabilities of neural BP decoders. \\ \hline

        \multirow{4}{*}{Other Approaches} &
        Clausius et al. \cite{clausius2024graph} & 
        GNN-based joint equalization and decoding. \\  \cline{2-3}

        & 
        Wiesmayr et al. \cite{wiesmayr2022duidd} & 
        Deep-unfolded interleaved detection and decoding for MIMO wireless systems. \\ \cline{2-3}

        & 
        Lee et al. \cite{lee2021neural} & 
        Learning-aided multi-round BP decoding with impulsive perturbation. \\ \cline{2-3}

        & 
        Wang et al. \cite{wang2019low} & 
        DL detection of decodable codewords for reducing the decoding delay. \\ \hline
    \end{tabular}
    \label{tab:bp}
\end{table*}

\begin{figure}[t]
     \centering
     \begin{subfigure}[b]{1.0\linewidth}
         \centering
         \includegraphics[width=0.9\linewidth]{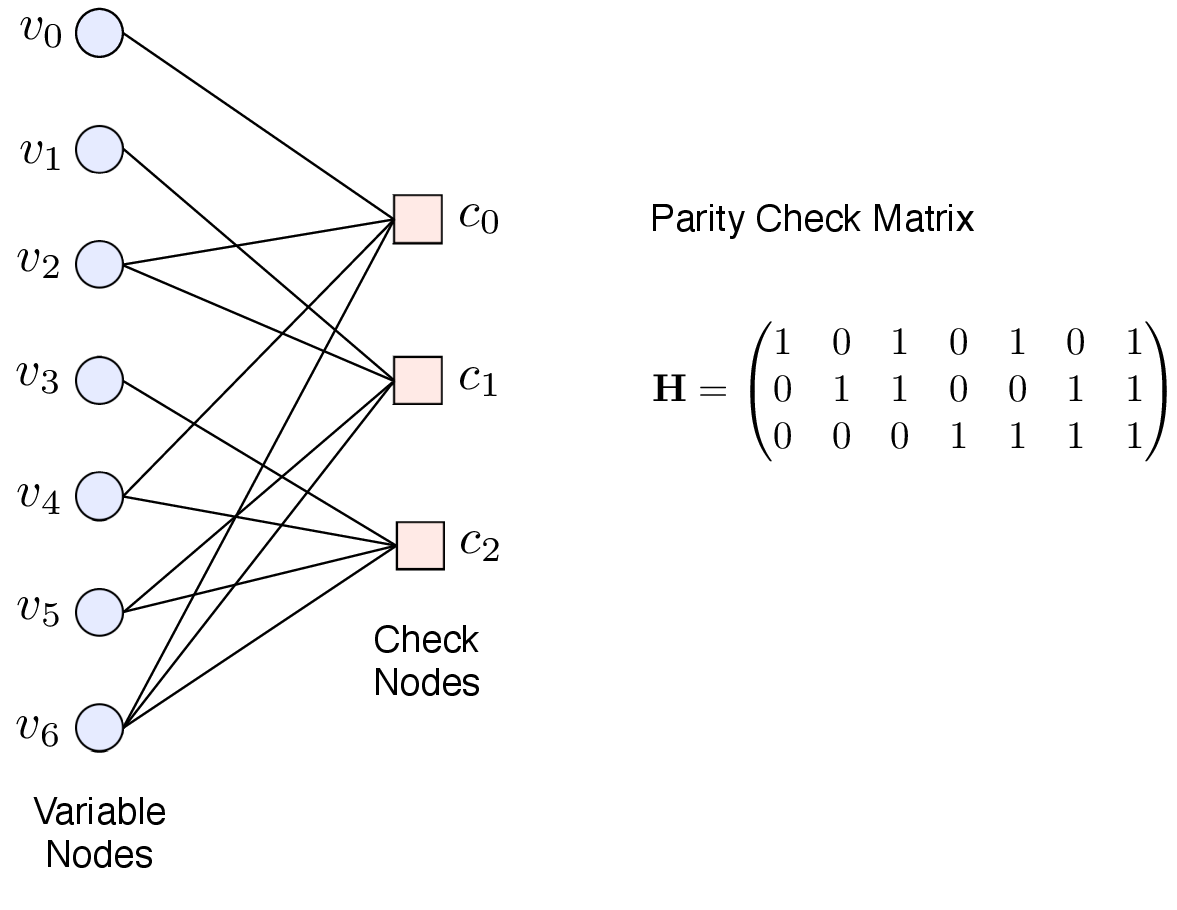}
         \caption{PCM and corresponding Tanner graph.}
         \label{fig:bp_pcm}
     \end{subfigure}
     \\ \vspace{5mm}
     \begin{subfigure}[b]{1.0\linewidth}
         \centering
         \includegraphics[width=0.95\linewidth]{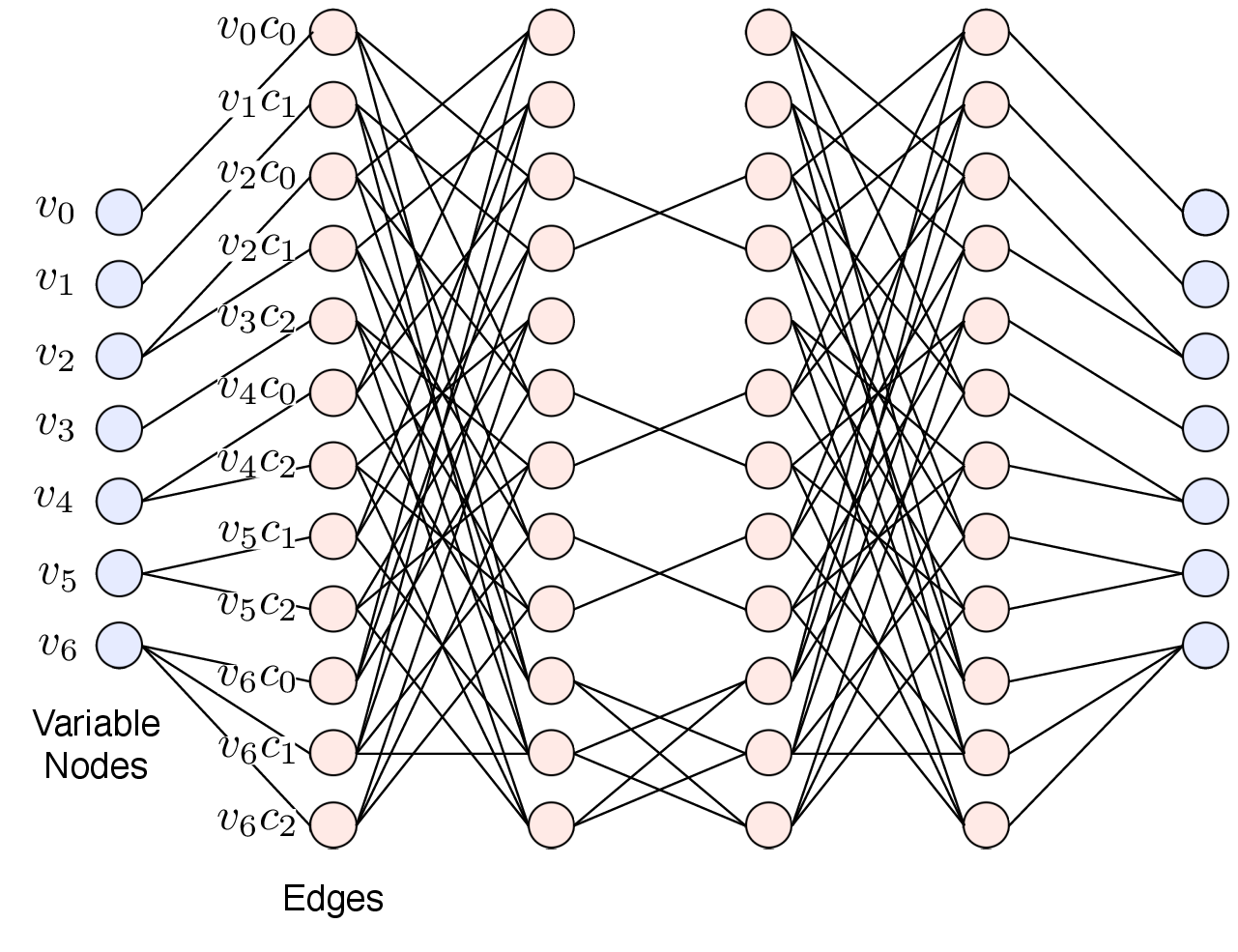}
         \caption{Unrolled BP trellis (two iterations).}
         \label{fig:bp_unrolled}
     \end{subfigure}
    \caption{An example of deep unfolded BP decoder for (7, 4) Hamming codes.}
    \label{fig:bp}
\end{figure}

BP decoding is an efficient iterative decoding algorithm that is commonly used for decoding LDPC codes. BP decoding is performed on Tanner graph consisting of CNs and VNs, which correspond to codeword bits and parity check equations, respectively. An example of PCM with the corresponding Tanner graph representation of a (7, 4) Hamming code is shown in Fig.~\ref{fig:bp_pcm}.
In BP decoding, decoding messages are iteratively updated at CNs and VNs based on the Bayes' rule. In practice, the min-sum approximation \cite{fossorier1999reduced} is applied to the CN updates to reduce complexity, and this decoding is referred to as \ac{ms} decoding. The performance loss due to the min-sum approximation can be compensated for by \ac{nms} and \ac{oms} decoders \cite{chen2005reduced} at the cost of slightly increased complexity.

In \cite{nachmani2016learning}, the authors proposed a DL-based implementation of BP decoding by treating BP decoding as a differentiable process, where the decoding messages are passed through unrolled iterations in a feed-forward fashion. Additionally, trainable weights were introduced at the edges, which are then optimized via SGD.
An example of the unrolled BP trellis for a (7, 4) Hamming code is illustrated in Fig.~\ref{fig:bp_unrolled}. For code length $N$, the number of edges $E$, and the number of iterations $L$, the unfolded trellis has $N$ neurons at the input and output layers, and $E$ neurons at the $2L$ hidden layers. The network architecture is a non-fully connected neural network.
As we review below, the trainable BP decoder over the unfolded trellis has been extensively studied in the literature. We summarize these works in Table~\ref{tab:bp}.

Note that the idea of unfolding an iterative algorithm into a structure analogous to a neural network, i.e., deep unfolding \cite{hershey2014unfolding}, is a common model-based DL approach \cite{shlezinger2023model} often considered in the design of communication systems. Besides channel decoding, the idea of deep unfolding has been successfully applied, for example, to MIMO signal detection and channel estimation \cite{he2019model,balatsoukas2019deep,jagannath2021redefining}.

\subsubsection{Neural MS Decoders and Its Variants}

As mentioned above, the paper \cite{nachmani2016learning} was the initial work that applied a feedforward network to BP decoding, where trainable weights are assigned to the edges of the factor graph, which are then optimized via SGD over the unrolled iterative BP decoding. The proposed parameterized decoder can compensate for the effect of small cycles in the Tanner graph of the code by properly scaling the weights. The effectiveness of the proposed decoder has been demonstrated for short BCH codes, for which standard BP decoding does not work well due to many short cycles in the graph. Later, in \cite{nachmani2017rnn}, the same authors extended the work by introducing an RNN architecture and showed that this architecture leads to comparable performance to the feedforward architecture in \cite{nachmani2016learning} even with fewer parameters. In \cite{nachmani2018deep}, the authors also proposed neural network-based NMS and OMS decoders for reducing the complexity of the BP decoder, which are the generalized versions of the standard NMS and OMS decoders \cite{chen2005reduced}. Neural network-based OMS decoding has also been studied independently in \cite{lugosch2017neural}.
Another approach to enhance the performance of MS decoding while preserving the low complexity property was considered in \cite{wu2018decoding}; the authors proposed \ac{lams} for \ac{pbldpc} codes, where the magnitudes of the check node updating and channel output are linearly adjusted by a small and shallow neural network.
In contrast to the above-mentioned studies that considered the flooding schedule, the authors in \cite{shah2021neural} proposed a neural network-aided \ac{noms} decoding for the layered BP with application to 5G LDPC codes.

As another variant of MS decoding, the \ac{ams} decoding proposed by Qualcomm \cite{richardson2018adjusted} has drawn attention, where it employs \acp{lut} to simplify nonlinear CN processing.
In \cite{yu2023neural}, the authors introduced a neural network based selection mechanism to AMS decoding that automatically selects the check node updating rule from either the MS rule or the BP rule and demonstrated that the proposed decoder outperforms the conventional neural NMS decoder. The \ac{smms} algorithm \cite{darabiha2006bit} is another variant of MS decoding that simplifies the CN update in the MS algorithm. Specifically, in SMMS decoding, only one minimum magnitude is calculated at each CN over all the CN inputs and a correction is applied to outgoing messages if required. The SMMS decoder can be further improved by the \ac{vwms} algorithm \cite{angarita2014reduced}, which introduces variable correction factors into the CN update that depend on the number of iterations. To efficiently learn the optimal correction factors in the VWMS algorithm, the authors in \cite{hsu2024gc} proposed a neural network-aided approach, instead of an exhaustive search in the original work \cite{angarita2014reduced}. The effectiveness of the proposed scheme in terms of throughput was demonstrated experimentally using the 40~nm CMOS TSMC process.
Unlike the above-mentioned methods, which simplify CN updates, \ac{scms} decoding \cite{savin2008self} modifies the VN processing by deleting unreliable messages. More specifically, in SCMS decoding, any variable node message that changes sign between two consecutive iterations is discarded, i.e., set to zero. The authors in \cite{kim2024neural} introduced trainable normalization and offset weights to the SCMS decoder, which are trained by DL techniques. It was demonstrated that the error rate performance of the proposed neural SCMS decoder is close to that of the BP decoding.

Although the above-mentioned works applied neural BP decoding to LDPC codes, it has also been used to decode polar codes. In \cite{xu2017improved}, similar to \cite{nachmani2016learning}, a neural MS decoder was applied to factor graphs of polar codes. The proposed decoder was demonstrated by simulations to outperform conventional BP decoding with the same number of iterations. Also, the authors presented an efficient hardware implementation of the basic computation block of the proposed decoder. In \cite{xu2018polar}, a similar trainable BP decoding was applied to sparse graphs of polar codes \cite{cammerer2018sparse}, which was shown to achieve comparable performance to BP decoding even with a single trainable parameter. Furthermore, the authors in \cite{dai2018new} proposed an NOMS decoder for polar codes that introduces both normalization (or scaling) factors and offsets, and demonstrated that the proposed decoder achieves better performance than the state-of-the-art schemes, including the decoder proposed in \cite{xu2017improved}.

In order to enhance the performance of a standalone polar code and close the performance gap from CA-SCL decoding with lower latency, the authors in \cite{doan2019crc} proposed neural BP decoding for polar codes concatenated with a CRC code by exploiting the concatenated factor graph of the polar code and CRC, while the conventional BP decoding for concatenated CRC-polar codes is applied only to the factor graph of polar codes and the CRC is used only to verify the result of BP at each iteration.
Furthermore, the authors in\cite{xu2020deep} considered concatenated polar and LDPC codes and proposed two-dimensional OMS decoding. They optimized the trainable parameters of the decoder by back propagation over the unfolded BP trellis and showed that the performance of the proposed decoder is comparable to the exact BP decoder.

\subsubsection{Performance Enhancement}
To enhance BP decoding of polar codes, the authors in \cite{teng2019convolutional,teng2020convolutional,teng2021convolutional} proposed to combine BP decoding with a CNN-assisted bit flipping mechanism, which performs the flipping bit selection in the BP-BF decoder \cite{yu2019belief} based on a CNN trained using the metadata of the BP decoding process. The authors demonstrated that the proposed scheme can achieve a lower BLER than SCL decoding. Meanwhile, in order to reduce the computational complexity of the CNN-based approach, the paper \cite{sun2020lstm} proposed an LSTM network that predicts error-prone bits to be flipped based on the magnitude of \ac{llr} after the original BP decoding.

The authors in \cite{nachmani2019hyper} introduced a hypernetwork \cite{chauhan2023brief} that generates weights of a neural BP decoder to make the decoder more adaptive by letting the weights be a function of the node's input. The same authors also introduced hypernetworks for decoding short polar codes \cite{nachmani2020gated} and showed that the proposed decoder achieves similar BER as SCL decoding in the high SNR region. Furthermore, they proposed an autoregressive BP decoder that incorporates the estimated SNR and multiple autoregressive signals obtained from the intermediate output of the network \cite{nachmani2021autoregressive}.

Training data preparation is an essential part of training DNN-based decoders. In particular, the choice of training SNR plays an important role in training a DL-based channel decoder for generalization. A common approach is to train the decoder over varying SNR ranges \cite{nachmani2016learning}. Besides, the optimal choice of the training SNR has been studied either empirically \cite{kim2018communication,gruber2017deep} or analytically \cite{benammar2018optimal}. To address the problem of choosing the optimal training SNR, the authors in \cite{be2019active} proposed \emph{active deep decoding}, inspired by active learning \cite{settles2009active}. Specifically, based on the observation that no optimal training SNR for all validation sets exists, the paper proposed to adaptively sample training data instead of passively generating examples during training. It was demonstrated that this active deep decoding scheme offers performance gain by effectively sampling the training data without increasing the inference (decoding) complexity.

In \cite{raviv2020data}, inspired by ensemble models that are widely used to solve complex tasks by decomposing them into multiple simpler tasks each of which is solved locally by a single expert member of the ensemble \cite{rokach2010ensemble}, the authors introduced the ensemble of neural BP decoders. The proposed scheme consists of a single classical \ac{hdd} and multiple trainable BP decoders, where the classical HDD is employed to assign a received codeword to a single expert BP decoder based on the number of the estimated codeword errors. It was demonstrated that this scheme achieves remarkable performance gains over the single neural BP. Furthermore, the data-driven ensemble scheme has been extended to BP polar decoders in \cite{raviv2023crc,raviv2024crc}.

Meanwhile, many practical LDPC codes exhibit an error floor\footnote{For modern iteratively decodable codes, such as LDPC codes and turbo codes, there is an SNR point after which the error rate decreases only slowly. This phenomenon is called \emph{error floor}.}. For applications such as ultra-reliable and low-latency communications that require extremely low BLER, it can be critical to mitigate the error floor. Since the error floor of LDPC codes is commonly attributed to the suboptimality of the iterative message passing decoding algorithms for factor graphs with cycles, the paper \cite{kwak2023boosting} proposed training methods for neural NMS decoders to eliminate the error floor of LDPC codes. Specifically, inspired by the boosting learning technique \cite{freund1997decision}, the authors divided the decoder into two cascaded neural decoders and trained the first decoder to improve the waterfall performance, while the second decoder was trained to handle the residual errors that are not corrected by the first decoder.

\subsubsection{Variants of Random Redundant Decoding}
The \ac{rrd} algorithm \cite{halford2006random} and \ac{mbbp} decoder \cite{hehn2007multiple} are other approaches to soft-decision decoding of short block codes based on a redundant PCM.
\ac{mrrd} \cite{dimnik2009improved} is an algorithm that attempts to benefit from both RRD and MBBP decoding, which make use of a permutation group (automorphism group) of codes and parallel iterative decoders, respectively.

In \cite{liu2020deep}, the authors proposed a \ac{ncrd} algorithm for \ac{hdpc} codes in order to improve the performance of RRD decoding. The NC-RD algorithm introduces two preprocessing steps to the RRD decoding. More specifically, the algorithm first classifies the variable nodes of the parity-check matrix by the $k$-median algorithm based on the number of shortest cycles associated with each variable node, and then generates a list of permutations of bit positions from the automorphism group based on the permutation reliability metrics. The authors further proposed the neural network-based NC-RD algorithm by unfolding the NC-RD decoding process and introducing trainable weights.

\begin{figure}[t]
\centering
\includegraphics[width=0.7\linewidth]{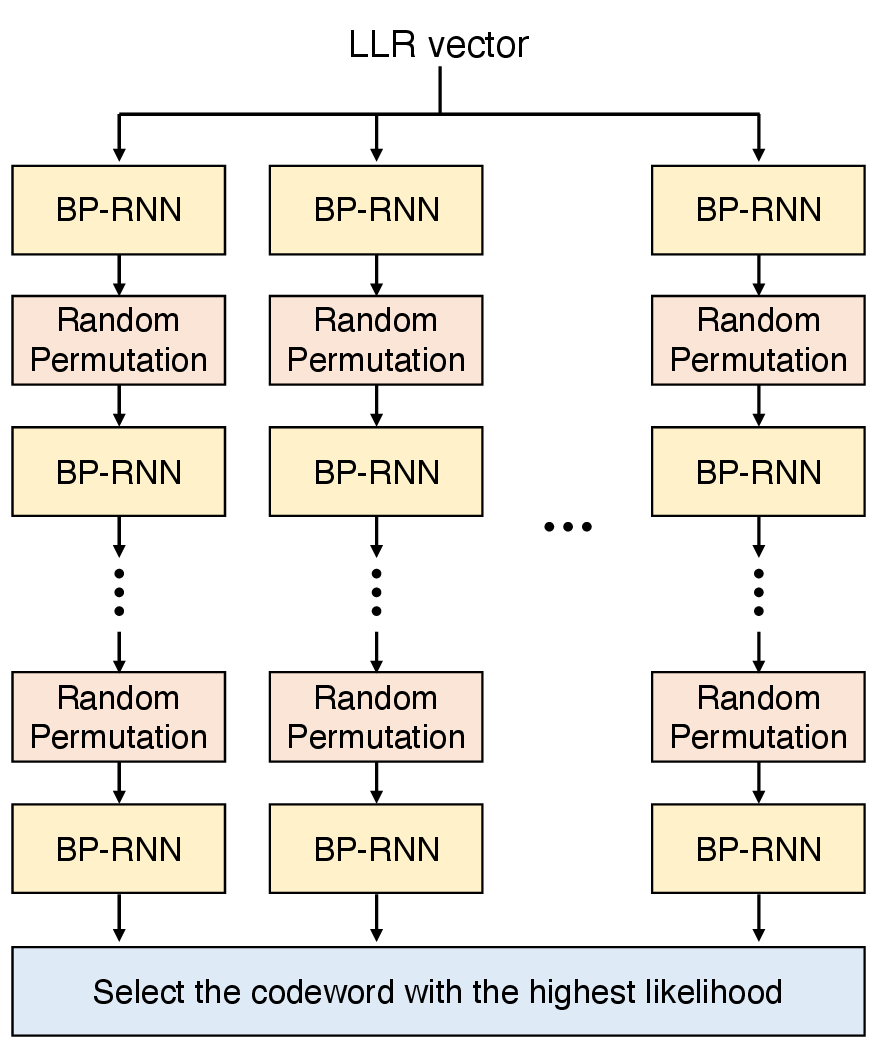}
\caption{The mRRD algorithm with RNN-based BP decoders in \cite{nachmani2018deep}.}
\label{fig:mRRD} 
\end{figure}

In \cite{nachmani2018deep}, the authors applied the concept of DL-based BP decoding to mRRD \cite{dimnik2009improved} by replacing the BP decoding blocks in the mRRD algorithm with the proposed RNN-based BP decoders. The resulting decoder structure is shown in Fig.~\ref{fig:mRRD}. The proposed RNN-based mRRD decoder has been demonstrated to achieve near maximum-likelihood performance with reasonable computational complexity.

\subsubsection{Optimization-Based Decoding}
Optimization-based decoding is another research direction aimed at improving the performance of BP decoders. The origin of the optimization-based decoding goes back to the work by Feldman who introduced a \ac{lp} formulation of decoding LDPC codes \cite{feldman2003decoding}.

The LP decoder is based on the LP relaxation of the original maximum-likelihood decoding problem \cite{candes2005decoding}. However, the LP decoder has higher computational complexity and worse error-correcting performance in the low SNR region compared with the BP decoder. In order to address the above drawbacks, the paper \cite{wei2020admm} proposed a trainable \ac{admm}-penalized decoder \cite{barman2013decomposition} by unfolding the ADMM iterations. It was demonstrated that the proposed decoder can outperform the conventional BP decoder in high SNR region with comparable execution time.

Afterwords, the authors in \cite{wadayama2019deep} introduced a trainable projected gradient decoder for LDPC codes by unfolding the PGD algorithm and optimizing the parameters via backpropagation. The proposed decoder alternately performs the gradient and projection steps, where the former moves in the direction of the negative gradient of the objective function, while the latter maps the search point into a feasible region that nearly satisfies the optimization constraint. 
Also, in \cite{wadayama2023proximal}, the same authors proposed proximal decoding of LDPC codes based on the proximal gradient method \cite{parikh2014proximal}, which is used for solving non-differentiable convex optimization problems.

\subsubsection{RL-Based Approaches}
It is known that parallel BP decoders on independently permuted factor graphs can significantly improve the performance of single BP decoding for polar codes \cite{elkelesh2018belief,doan2018decoding}. In \cite{doan2020decoding}, the authors addressed the problem of selecting the permutations on the factor graph that lead to successful decoding given a channel observation. Specifically, they viewed the selection of permutations as a multi-armed bandit problem and proposed an RL-based CRC-aided BP decoder that attempts to select the best set of permutations. The proposed scheme was shown to achieve better performance than other approaches such as cyclically-shifted and random
factor-graph permutations \cite{hussami2009performance,elkelesh2018belief}.

In \cite{habib2020learned,habib2021belief,habib2023reldec}, the authors proposed a novel RL-based sequential BP decoding scheme to optimize the scheduling of CN clusters for moderate length LDPC codes\footnote{In contrast to the standard flooding scheduling where all CNs and VNs are updated simultaneously at each iteration, sequential BP decoding, or layered decoding, updates nodes or sets of nodes individually in sequence.}. In the proposed scheme, $m$ CNs are divided into sets of $z$ CNs, called \emph{cluster}, and the scheduling problem, i.e., cluster selection with $\lceil m/z \rceil$ possible actions, was optimized by Q-learning. Furthermore, they proposed novel meta-learning based sequential decoding schemes to quickly adapt to changing channel conditions due to fading in wireless scenarios. The RL-based scheduling of sequential BP decoding has also been proposed for generalized LDPC codes \cite{habib2023reinforcement}, where the authors showed that the proposed RL-based decoding scheme was shown to significantly outperform the standard BP flooding decoder, as well as a sequential decoder based on random scheduling with the smaller number of CN updates.

\subsubsection{Customized Loss Function}
As in Section~\ref{sec:synd}, a syndrome loss function has been used for DL-based BP decoders. In \cite{lugosch2018learning}, the authors proposed a soft syndrome as a loss function for training a neural BP decoder. Unlike the paper \cite{bennatan2018deep}, which utilizes the hard syndrome as an input to the decoder, this paper introduced the \emph{soft} syndrome which is defined similarly to the CN update rule in MS decoding, in addition to the conventional cross entropy loss function. 

While the application of \cite{lugosch2018learning} was limited to decoders that output a soft estimate of the codeword, this is not the case for polar decoders that do not use a PCM. To address this, the authors in \cite{teng2020syndrome} proposed two modified syndrome losses: frozen-bit syndrome loss and CRC-enabled syndrome loss. The authors also introduced a syndrome-enabled blind equalizer based on the proposed syndrome loss, which does not require the transmission of training sequences.

Unlike the above syndrome-based approaches, the authors in \cite{nachmani2022neural} considered a linear combination of sparse node loss and knowledge distillation loss, in addition to the conventional cross entropy loss. Knowledge distillation is a technique in DL where one uses a teacher network to guide the training of a smaller student neural network \cite{hinton2015distilling}.
Sparse node loss imposes a sparse constraint on the node activations based on the $L_p$ norm, whereas the knowledge distillation loss aims to mimic the teacher network, which was the standard MS decoder without trainable parameters, by transferring knowledge. It was shown that the proposed loss terms provide BER performance improvement of up to $1.1$~dB without increasing the runtime complexity and the model size.

\subsubsection{Memory and Complexity Reduction}
\label{sec:memory}
\begin{figure}
     \centering
     \begin{subfigure}[b]{1.0\linewidth}
         \centering
         \includegraphics[width=0.95\linewidth]{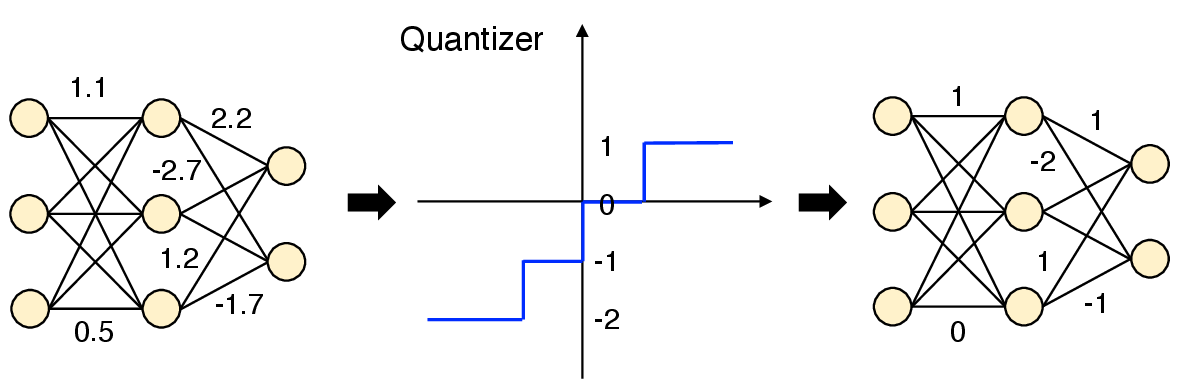}
         \caption{Quantization of weights.}
         \label{fig:quant}
     \end{subfigure}
     \\ \vspace{5mm}
     \begin{subfigure}[b]{0.45\linewidth}
         \centering
         \includegraphics[width=0.6\linewidth]{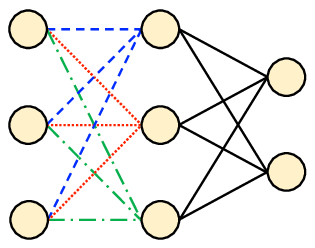}
         \caption{Sharing weights (connections with the same color have the same weight).}
         \label{fig:share}
     \end{subfigure}    
     \hspace{3mm}
     \begin{subfigure}[b]{0.45\linewidth}
         \centering
         \includegraphics[width=0.6\linewidth]{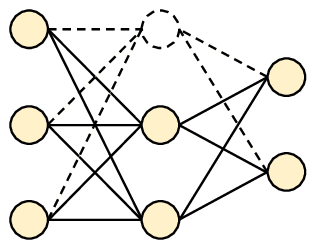}
         \caption{Pruning neurons (pruned neurons are indicated by dashed circles).}
         \label{fig:prune}
     \end{subfigure}
    \caption{Approaches to reducing computational and space complexities of a DL decoder.}
    \label{fig:complexity}
\end{figure}

A neural network-based BP decoder introduces different weights (or scaling factors) to different edges in the Tanner graph, which can significantly increase the computational and space complexities of standard BP decoding. In fact, this issue has been studied extensively for generic DNNs, i.e., not limited to channel decoding applications, due to the large parameter sizes of modern DL models \cite{thompson2020computational}. A popular approach is neural network pruning \cite{karnin1990simple,han2015deep,liu2018rethinking,blalock2020state,hoefler2021sparsity}, which aims to remove redundant parameters of an original network while preserving the accuracy. Parameter shaping is also an effective way to reduce parameters by sharing parameters between different neurons. Parameter shaping is typically exploited in CNN, where all neurons in a particular feature map share the same weight. Another popular approach is parameter quantization \cite{nagel2021white,gholami2022survey}. These major approaches to addressing the complexity problem are illustrated in Fig.~\ref{fig:complexity}.

\textit{(a) Quantization:}
In \cite{teng2019low}, the authors proposed a weight quantization mechanism for an RNN polar decoder. Specifically, they proposed a two-step approach, where floating-point weights are quantized into $2^q$ quantization levels and then they are further compressed into $2^c$ ($c<q$, $c,q \in \mathbb{N}$) quantization levels, which are the most commonly used among $2^q$ quantization levels.
The quantization of the RNN polar decoder was also studied in \cite{ibrahim2022enhanced}, where the authors demonstrated that quantization after training leads to better performance compared to the case where quantization is applied after every epoch during training.

Instead of quantizing DNN decoder parameters, several papers have investigated quantizing decoding messages or LLR values computed from channel observations. For example, in \cite{wadayama2018joint}, the authors trained a parameterized quantization of LLR values that maximizes the performance of BP decoding. Similarly, the authors in \cite{geiselhart2022learning} investigated the design of quantizers in an LDPC decoder that are used for quantizing both LLRs and iterative decoding messages. On the other hand, the authors in \cite{gao2020learning} proposed to train neural BP decoder for the system with one-bit quantizer.

Unlike existing studies that consider the AWGN channel, the authors in \cite{vasic2018learning,xiao2019finite,xiao2020designing} considered BSC and proposed finite precision decoders, called \ac{faid}, for LDPC codes with \ac{rqnn}. More specifically, they proposed the BER as the loss function to train the RQNNs over BSC by leveraging \ac{ste} \cite{bengio2013estimating} to overcome the issue of gradients vanishing caused by the low precision activations in the RQNN and quantization in the BER.

In \cite{lyu2023optimized}, the authors proposed a joint optimization of  quantized message alphabets and quantization thresholds. Specifically, the authors utilized the softmax distribution \cite{wang2022learnable} to make the quantization thresholds trainable by softening the one-hot distribution of the quantization. The proposed decoder was shown to outperform the original non-surjective FAIDs \cite{nguyen2016non} in terms of error rate performance.

\textit{(b) Weight Sharing:}
In \cite{lian2018can,lian2019learned}, the authors proposed simple-scaling models for weighted BP decoding in \cite{nachmani2016learning} that share weights across edges, using only three scalar parameters per iteration: message weight, channel weight, and damping factor. The authors showed that such simple scaling models are often sufficient to achieve gains similar to the fully parameterized decoder. 

The authors in \cite{wang2020model,wang2022normalized} proposed a parameter sharing scheme for a neural NMS decoder that shares the same correction (normalization) factors in the same layer. In contrast, the authors in \cite{wang2021neural,wang2023ldpc} proposed a family of weight sharing schemes for a neural NMS decoder that uses the same weight for edges with the same check node degree and/or variable node degree.
Similarly, the authors in \cite{dai2021learning} proposed a neural MS decoder for protograph LDPC codes where a bundle of edges derived from the same edge type share identical parameters. Due to the lifting structure of protograph LDPC codes, the same set of parameters can be employed for multiple codes derived from the same base code. 

In \cite{liang2022low}, the authors applied the \ac{tt} decomposition \cite{oseledets2011tensor} to a neural NMS decoder, where it decomposes a high-order tensor into  several low-order tensors.
This not only reduces the number of weight parameters, but also the number of multiplications required in the CN and VN updates. Furthermore, in \cite{liang2023joint}, the same authors proposed \ac{tr} decomposition \cite{zhao2016tensor} combined with weight sharing to further reduce the storage and computational complexity.

In \cite{cheng2023rate}, a weight sharing scheme was proposed for a neural BP or MS decoder to adapt to multiple code rates with a reasonable amount of parameters. Specifically, instead of training different decoders, they proposed to train a single rate-compatible decoder based on multi-task learning, where different parts of the parameters are activated to handle different code rates.

\textit{(c) Pruning:}
In \cite{buchberger2020pruning}, the authors proposed a novel pruning-based neural BP decoder for short linear block codes. The key idea was to prune unimportant CNs with small weights of an overcomplete PCM. Similarly, in \cite{buchberger2021learned}, the same authors proposed a neural BP with decimation \cite{filler2007binary} for LDPC codes. In particular, they identified the least reliable VN with the aid of DL, i.e., the VN with the lowest absolute \emph{a posteriori} LLR, and then decimated it to $\pm \infty$. It has been demonstrated that the proposed decoder with decimation can significantly outperform the conventional neural BP decoder.

\subsubsection{GNN Decoders}
In general, BP computes the optimal (posterior) marginal probability distributions only for a non-loopy graphical model, and in practice it often computes a poor approximation of the true distribution. To tackle this limitation, the authors in \cite{satorras2021neural} extended the standard GNN equations to factor graphs and presented a hybrid inference model that combines messages from BP and from GNN, where the GNN messages are learned to complement the BP messages. 

Instead of the method in \cite{satorras2021neural}, which extends BP decoding by combining it with a GNN, the authors in \cite{cammerer2022graph} proposed a fully GNN-based decoder.
In contrast to weighted BP decoding, they introduced two types of MPNN-based trainable message update functions: \emph{the edge message update functions} and \emph{the node update functions}.
Independently, an edge-weighted GNN decoder has been proposed in \cite{tian2023scalable}. In the proposed decoder, they applied an MPNN for updating messages and assigned a trainable weight to each edge message, which is optimized by a fully-connected feed-forward neural network, i.e., MLP. The major advantage of these GNN decoders over the standard neural BP decoder is that the number of trainable parameters is not affected by the code length. Therefore, after training, the trained decoder can be applied to codes with different rates and lengths without retraining.

\subsubsection{Understanding Neural BP Decoders}
In \cite{ankireddy2023interpreting}, the authors empirically showed how the learned weights mitigate the effect of short cycles in Tanner graphs to improve the reliability of the posterior LLRs and contribute to the robustness of the decoders across channels. The authors also introduced an analytical approach for finding the weights using GA and compared the neural MS decoders, showing that for complicated fading channels, the neural network-based weight optimization leads to better performance than the GA-based optimization.

The authors in \cite{adiga2024generalization} theoretically investigated the generalization capabilities \cite{mohri2018foundations} of neural BP decoders, i.e., the difference between empirical and expected BERs. The paper presented new theoretical results that bound the gap and showed its dependence on the decoder complexity, in terms of code parameters (such as message/code lengths, VN/CN degrees), decoding iterations, and the training dataset size. They empirically observed that the generalization gap increases with decoding iterations and code length, and decays with the training dataset size, supporting the theoretical results in their paper.

\subsubsection{Other Approaches}
To improve the decoding of short Raptor-like LDPC codes, the authors in \cite{lee2020multi} considered multi-round BP decoding with impulsive perturbation \cite{xiao2007perturbation}. Perturbation is a process of making a small intentional change in the received signal, and this scheme iteratively performs conventional BP decoding and perturbation until a valid codeword is found. In \cite{lee2021neural}, the authors proposed a neural network based perturbation symbol selection scheme where the symbols to be perturbed are selected from a pre-trained neural network and showed that the proposed scheme performs better than existing schemes such as \cite{lee2020multi} for Raptor-like LDPC codes. 

The performance of a standalone neural BP decoder could be further enhanced by jointly optimizing signal detection and decoding.
For instance, in \cite{wiesmayr2022duidd}, iterative signal detection and decoding via deep unfolding was proposed for MU-MIMO-OFDM.
In \cite{clausius2024graph}, the authors proposed GNN-based joint detection and decoding for \ac{isi} channels.

Besides the approaches introduced in Section~\ref{sec:memory}, the authors in \cite{wang2019low} proposed another approach to alleviate decoding complexity and latency. Specifically, they proposed a DL approach for detecting the decodable codewords and predicting the iteration number from the received signal to reduce the decoding delay. This could potentially be useful for early feedback prediction in \ac{harq}. Furthermore, in \cite{han2024accelerating}, the authors accelerated neural BP decoding through coded distributed computing \cite{li2017fundamental}. In particular, they reformulated the neural BP decoding operations as matrix-vectors to facilitate distributed parallel decoding. 

In addition to communication systems, DL-based decoders have also been studied for storage systems.
In \cite{zhong2023deep}, the authors proposed DL-based decoder for \ac{sttmram} \cite{chen2010advances}. In order to adapt to the process variation and unknown offset of the resistance caused by the change in working temperature, the authors proposed an adaptive decoding scheme based on the three DNN decoders, i.e., BF, MS, and BP decoders, which share the same DNN architecture but have different weights. 
In \cite{zhong2020deep}, the same research group proposed a neural normalized offset \ac{rbms} decoding algorithm for STT-MRAM by introducing trainable parameters to the RBMS algorithm \cite{chen2011comparisons}. 
It has been demonstrated that the proposed scheme can outperform the RBMS algorithm over the STT-MRAM channel, while maintaining similar decoder structure and time complexity of the standard RBMS decoder.

Neural network-based BP decoding has been studied not only for classical error-correcting codes, but also for quantum error-correcting codes.
For example, neural BP decoding has been applied to quantum LDPC codes \cite{liu2019neural} for which standard BP decoding may be insufficient due to the error degeneracy feature of quantum error-correcting codes \cite{poulin2008iterative}. By designing the loss function to account for error degeneracy, the decoding accuracy was improved up to three orders of magnitude compared to the standard BP decoder without training. Neural BP decoding for quantum LDPC codes was also studied in \cite{gong2023graph,miao2023neural}.

\subsection{DL-Aided Decoding of Polar Codes}
\label{sec:PolarDecoder}

\begin{table*}[h]
    \centering
    \caption{Summary of DL-aided polar decoders.}
    \begin{tabular}{|p{4cm}|p{3.3cm}|p{9cm}|} \hline
        Category & Reference & Main Contribution \\ \hline

        \multirow{4}{*}{\shortstack{Neural Network-Based \\SC Decoding}} &
        Cammerer et al. \cite{cammerer2017scaling,gao2018neural} & 
        Partitioned neural network decoders. \\ \cline{2-3}

        &
        Doan et al. \cite{doan2018neural} & 
        Neural successive cancellation (NSC) decoding. \\ \cline{2-3}

        &
        Wodiany et al. \cite{wodiany2019low} & 
        Efficient implementation of an low-precision NSC decoder. \\ \cline{2-3}
               
        &
        Hebbar et al. \cite{hebbar2023crisp} & 
        Novel curriculum learning-based sequential neural decoder. \\ \hline
        
        \multirow{5}{*}{DL-Aided SCF Decoding} &
        Wang et al. \cite{wang2019learning} & 
        LSTM network that estimates the first erroneous bit. \\ \cline{2-3}
                       
        &
        He et al. \cite{he2020machine} & 
        LSTM-based identification of erroneous bits for DSCF decoding. \\ \cline{2-3}
        
        &
        Doan et al. \cite{doan2019neural,doan2021neural} & 
        Neural DSCF with trainable bit-flipping metric. \\ \cline{2-3}
        
        &
        Wang et al. \cite{wang2021reinforcement} & 
        Q-learning-assisted SCF decoding algorithm. \\ \cline{2-3}

        &
        Doan et al. \cite{doan2021fast} & 
        RL-based bit-flipping strategy for fast SC decoding. \\ \hline
        
        \multirow{6}{*}{DL-Aided SCLF Decoding} & 
        Hashemi et al. \cite{hashemi2019deep} & 
        Trainable bit-flipping metric for SCL decoding. \\ \cline{2-3}

        &
        Doan et al. \cite{doan2022fast} & 
        FSCLF decoding algorithm. \\ \cline{2-3}
 
        &
        Chen et al. \cite{chen2020low} & 
        LSTM-assisted bit-flipping algorithm for a CA-SCL decoder. \\ \cline{2-3}
        
        &
        Tao et al. \cite{tao2021dnc} & 
        New flip algorithm based on DNC. \\ \cline{2-3}
        
        &
        Liang et al. \cite{liang2023deep} & 
        Stacked LSTM to improve the accuracy of erroneous bit prediction. \\ \cline{2-3}
        
        &
        Li et al. \cite{li2024deep} & 
        Approximated bit-flipping metric for DSCLF decoding. \\ \hline
       
        \multirow{2}{*}{Other Approaches} &
        Lu et al. \cite{lu2023deep} & 
        DL-aided shifting metric for SCL decoding. \\ \cline{2-3}

        &
        Liu et al. \cite{liu2019exploiting} & 
        CRC error-correction aided SCL decoding. \\ \hline
        
    \end{tabular}
    \label{tab:polar}
\end{table*}

Model-free decoders in Section~\ref{sec:model-free} as well as neural BP decoders in Section~\ref{sec:BP} are easily applicable to polar codes. In the following, we also review methods for designing model-free decoders that take the specific code structure into account. Furthermore, we focus on DL approaches that augment conventional SC or SCL decoders, instead of replacing them with a DNN. In Table~\ref{tab:polar}, we provide the summary of these methods.

\subsubsection{Neural Network-Based SC Decoding}
Although the straightforward application of DNN is a viable option for decoding polar codes as in \cite{gruber2017deep}, the major issue was the exponential growth of training complexity.
In \cite{cammerer2017scaling}, the authors addressed this issue by introducing \ac{pnn} decoders. More specifically, inspired by the simplified successive cancellation algorithm in \cite{sarkis2015fast}, which divides the decoding tree into \acp{spc} and \acp{rc}), they replaced the SPC and RC subdecoders by neural networks. Simulations showed that the PNN decoder achieves similar BER performance to the SC and BP decoders for short lengths, such as $128$ bits, with potentially much lower latency. A similar concept of partitioning a neural network-based decoder for polar codes was also investigated in \cite{gao2018neural}. Furthermore, in order to reduce the latency of PNN, \ac{nsc} decoding of polar codes was proposed in \cite{doan2018neural}, where multiple constituent neural network decoders are incorporated into SC decoding, and its efficient implementation based on a low-precision neural network decoder was studied in \cite{wodiany2019low}.

Recently, another approach to tackle the difficulty of learning to decode long polar codes was proposed in \cite{hebbar2023crisp}, where a novel curriculum learning-based sequential neural decoder for polar and PAC codes was proposed. The paper designed a novel curriculum to train RNN, where the problem of joint estimation of information bits is decomposed into a sequence of sub-problems of increasing difficulty. The proposed decoder was shown to achieve better BER performance than the conventional supervised training without curriculum and standard SC decoding.

Instead of completely replacing a conventional decoder with DNNs, DL-based approaches that support conventional decoders such as SC and SCL decoding have been extensively studied. Among them, DL-assisted \ac{scf} decoding \cite{afisiadis2014low} is one of the most popular approaches, which will be reviewed in the following.

\subsubsection{DL-Aided SC Flip Decoding}
\begin{figure}[t]
\centering
\includegraphics[width=0.7\linewidth]{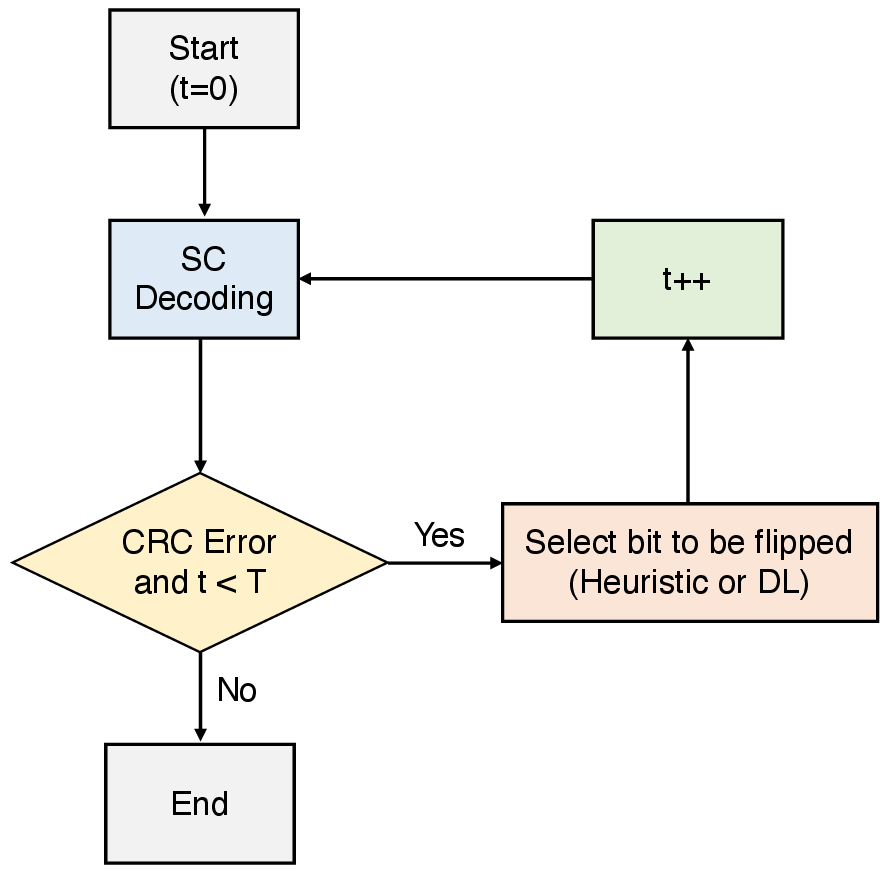}
\caption{Flowchart of an SCF decoding framework with the number of trials $T$.}
\label{fig:scf} 
\end{figure}

Despite its low-complexity, the error correcting performance of SC decoding at finite block lengths is not comparable to other modern codes such as LDPC codes. In order to improve the finite block length performance, SCF decoding has been proposed in \cite{afisiadis2014low} inspired by the fact that the first erroneous bit decision in SC decoding has a detrimental impact on the resulting error rate. The SCF decoder first performs standard SC decoding to generate a first estimated codeword, and if the codeword passes the CRC, decoding is complete. If the CRC check fails, the SCF decoding makes $T$ additional attempts to identify the first error in the codeword. In each attempt, a single estimated codeword bit is flipped with respect to the initial decision. The algorithm terminates when a valid codeword has been found or when all $T$ attempts have been made. The SCF decoding procedures is shown in Fig.~\ref{fig:scf}. SCF decoding retains the $O(N)$ memory complexity of the original SC algorithm and has an average computational complexity that is practically $O(N \log N)$ at high SNR, while still providing a significant gain in terms of error correcting performance. While SCF decoding is limited to correcting a single erroneous bit in the codeword, \ac{dscf} decoding \cite{chandesris2018dynamic,ercan2020practical} is a generalization of SCF-based decoding that is able to correct higher-order erroneous information bits by dynamically updating the set of flipping bit indices after each decoding attempt.

The common challenge in SCF and DSCF decoding is how to identify the first error bit that causes error propagation. In the original work \cite{afisiadis2014low}, the estimated codeword bits with the smallest amplitudes of LLR are flipped, but they are not necessarily the first errors. In fact, the optimal bit flipping strategy is still an open problem due to the lack of a rigorous mathematical characterization. Furthermore, DSCF decoding requires expensive exponential and logarithmic computations to compute the BF metric, which is used to determine the bit flipping position.

A popular DL-based solution is to train an LSTM network to estimate the first erroneous bit to be flipped. The authors in \cite{wang2019learning} have proposed an LSTM network for SCF decoding that takes an LLR sequence of the previous SC decoding attempt and outputs a vector where each element corresponds to the probability that a bit is the first error. Furthermore, the authors proposed a two-step training method that combines supervised learning with RL to train the LSTM to reverse previous incorrect flips. Similarly, the authors in \cite{he2020machine} proposed an LSTM-based error bit identification for DSCF decoding where the network is trained to identify the first erroneous bit and additional erroneous bits by supervised learning and RL, respectively.

There are other learning approaches that do not rely on model-free DNNs. In \cite{doan2019neural,doan2021neural}, the authors proposed neural DSCF decoding where they properly approximated and introduced trainable parameters to the BF metric and optimized its parameter by RMSProp, a variant of the SGD optimization technique. In \cite{wang2021reinforcement}, the authors proposed a Q-learning-assisted SCF decoding algorithm that selects the candidate flipping bits through the learned Q-table instead of metric sorting. It was demonstrated that the proposed decoding algorithm is particularly effective in reducing the decoding delay caused by sorting during the decoding process without sacrificing performance. Similarly, in \cite{doan2021fast}, an RL-based BF strategy is also investigated for fast SC decoding of polar codes \cite{alamdar2011simplified}, where the authors developed a new parameterized BF model based on \cite{hashemi2019deep} and optimized the trainable parameters using the policy gradient method.

\subsubsection{DL-Aided SCL Flip Decoding}
The DL-aided BF mechanism can be applied not only to SC decoding but also to SCL decoding to further improve the performance.

In \cite{hashemi2019deep}, the authors proposed the BF metric for SCL decoding, which is expressed by a trainable correlation matrix representing the likelihood of each decoded bit.
They optimized the trainable matrix using the RMSprop optimizer and demonstrated that compared to the conventional metric in \cite{chandesris2018dynamic}, the proposed BF metric significantly reduces computational complexity associated with the bit metric calculation while maintaining similar error rate performance.

In \cite{doan2022fast}, the authors applied BF to \ac{fscl} decoding \cite{hashemi2016fast,hashemi2017fast}, referred to as the \ac{fsclf} decoding algorithm, to address the high latency problem associated with the \ac{sclf} decoding algorithm. Specifically, the authors introduced a BF strategy tailored to FSCL decoding that avoids tree-traversal in the binary tree representation of SCLF to reduce the latency of the decoding process, and then derived a path selection error metric with a trainable parameter. The proposed decoder was shown to significantly reduce the average decoding latency, average complexity, and memory consumption of the SCLF decoder at the cost of slight degradation in error rate performance.

In contrast to the above approaches that do not utilize a DNN, several papers have proposed a neural network-based selection of the flipping bit position. For instance, in \cite{chen2020low}, an LSTM-assisted BF algorithm has been proposed for a CA-SCL decoder. Furthermore, the authors have used the domain knowledge to reduce the complexity and memory requirements and computational complexity for efficient hardware implementation.

The authors in \cite{tao2021dnc} proposed a new flip algorithm using \ac{dnc} \cite{graves2016hybrid}, which can be considered as an LSTM augmented with an external memory through attention-based soft read and write mechanisms. The proposed decoding algorithm is a two-phase decoding assisted by the two DNCs, i.e., flip DNC and flip-validate DNC. The former ranks flip positions for multi-bit flipping, while the latter is used to re-select flip positions when decoding fails. Simulation results show that the proposed DNC-aided SCLF achieves better error rate performance and reduction in the number of flipping attempts compared to the LSTM-based algorithms.

The authors in \cite{liang2023deep} proposed a stacked LSTM network to improve the accuracy of erroneous bit prediction. Specifically, they trained the three models separately: the first and second models predict the positions of the first and the second erroneous bits and the third model decides whether to continue flipping. Simulation results demonstrate that their proposed algorithms outperform existing SCLF decoding algorithms in terms of BLER performance and average number of decoding attempts.

In \cite{li2024deep}, the authors proposed an approximated error metric for \ac{dsclf} decoding of polar codes to improve the performance while keeping the average complexity low. To compensate for the approximation error, they introduced learnable parameters into the metric and optimized it through the custom neural network model using the RMSprop optimizer.

\subsubsection{Other Approaches to Enhancing SCL Decoding}
\Ac{sp} is another approach to enhance the performance of SCL decoding for polar codes \cite{rowshan2019improved}, which aims to prevent the correct path from being eliminated from the list. In \cite{rowshan2019improved}, it was demonstrated that the proposed SP mechanism offers remarkable performance gains over the BF approach. However, the SCL decoder with SP generally suffers from high computational complexity, due to the re-decoding attempts and the computation of the shifting metric. To alleviate this issue, the authors in \cite{lu2023deep} proposed a DL-aided shifting metric that is free from transcendental functions and can be computed on-the-fly based on the path metrics. 

Another approach to improve the performance of CA-SCL decoding was proposed in \cite{liu2019exploiting}, where the authors take advantage of the inherent error correction capability of CRC, i.e., not just for error detection. The authors performed CRC-based error correction using an LSTM network, where the LSTM network estimates the error pattern from the LLR sequence and the CRC syndrome. The proposed CRC error-correction aided SCL decoding scheme was demonstrated to outperform the error rate of the conventional CRC error-detection aided SCL decoding scheme at the same list size.

\subsection{DL-Aided Convolutional and Turbo Decoding}
\label{sec:TurboDecoder}
DL decoders have also been applied to convolutional and turbo codes. In particular, as we review below, several DL-aided turbo decoders have been proposed in recent years. We have listed these approaches in Table~\ref{tab:conv}.

\begin{table*}[h]
    \centering
    \caption{Summary of DL-aided convolutional and turbo decoders.}
    \begin{tabular}{|p{4cm}|p{3cm}|p{9cm}|} \hline
        Category & Reference & Main Contribution \\ \hline

        Convolutional Decoders &
        Teich et al. \cite{teich2020deep} & 
        A DNN decoder for convolutional codes. \\ \hline
        
        \multirow{4}{*}{Turbo Decoders} &
        Kim et al. \cite{kim2018communication} & 
        A neural network BCJR decoder, referred to as NEURALBCJR. \\ \cline{2-3}
                       
        &
        Jiang et al. \cite{jiang2019deepturbo} & 
        Deep turbo decoder (DEEPTURBO) trained in an end-to-end manner. \\ \cline{2-3}
        
        &
        He et al. \cite{he2020dnn} & 
        A novel model-driven decoder, called TurboNet. \\ \cline{2-3}
        
        &
        Hebbar et al. \cite{hebbar2022tinyturbo} & 
        TINYTURBO that reduces the parameters of TurboNet+. \\ \hline
        
    \end{tabular}
    \label{tab:conv}
\end{table*}

\subsubsection{Convolutional Decoder}
Convolutional codes encode the input stream by convolving with the generator polynomial, which can be efficiently implemented by shift registers. Convolutional codes can be represented by a time-invariant trellis, which allows efficient maximum-likelihood decoding based on the well-known Viterbi algorithm \cite{viterbi1967error}.

In order to improve the performance of \ac{itd} \cite{cardinal1999turbo}, the authors in \cite{teich2020deep} proposed a DNN decoder for convolutional codes by unfolding a \ac{hornn} decoder \cite{mostafa2012analysis}. The unfolded HORNN can be seen as a feedforward DNN whose parameters are trained by backpropagation with MBSGD. It was shown that with proper optimization of the parameters, the proposed decoder outperforms the conventional ITD and achieves performance close to maximum-likelihood decoding.

\subsubsection{Turbo Decoder}
Turbo codes, also known as parallel concatenated convolutional codes \cite{berrou1993near}, consist of two (usually identical) \ac{rsc} codes concatenated in parallel and a bit interleaver. Turbo codes are typically decoded by iterative decoding between two constituent SISO decoders where one constituent decoder computes posterior probabilities based on the \ac{bcjr} algorithm \cite{bahl1974284}, and then passes them to the other decoder. The turbo decoder significantly improves the error-correction performance by iteratively exchanging extrinsic information between the two constituent SISO decoders.

In \cite{kim2018communication}, the authors proposed a DL-based BCJR decoder for an RSC code based on bi-GRU, referred to as NEURALBCJR, and then extended it to a turbo decoder by replacing the component SISO decoder with the proposed NEURALBCJR decoder. Simulations demonstrated that the proposed scheme is particularly beneficial for non-Gaussian channels, such as $t$-distributed noise.
Later, a DL-aided turbo decoder, termed DEEPTURBO, was introduced in \cite{jiang2019deepturbo}. They also used bi-GRUs to replace the conventional SISO decoders as in \cite{kim2018communication}, but the authors in \cite{jiang2019deepturbo} trained different bi-GRU weights across different iterations, whereas the authors in \cite{kim2018communication} shared the same weight for all bi-GRU blocks. This enables a fully end-to-end training without imitating the BCJR algorithm. Furthermore, DEEPTURBO increased the number of posterior LLR values exchanged between the two decoders to expedite iterative decoding. Extensive simulations have demonstrated that DEEPTURBO exhibits an improved reliability, adaptivity, and lower error floor compared to NEURALBCJR. 
\begin{figure}
     \centering
     \begin{subfigure}[b]{1.0\linewidth}
         \centering
         \includegraphics[width=0.6\linewidth]{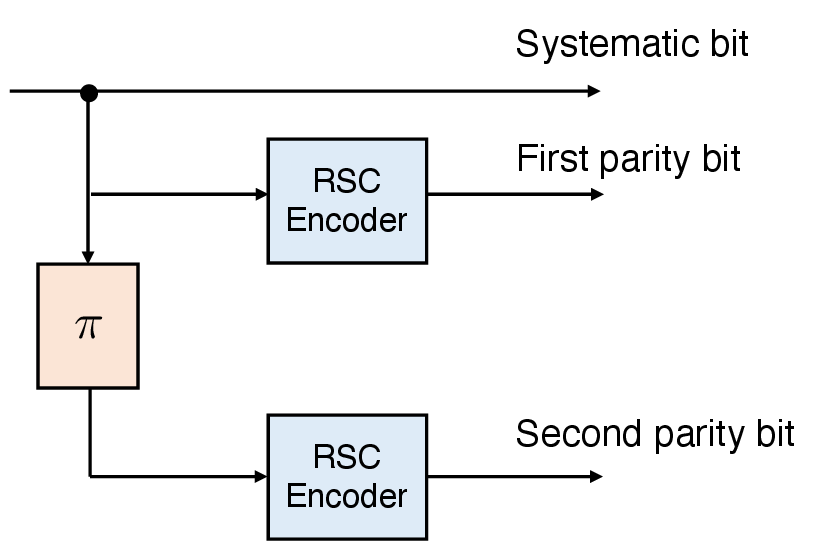}
         \caption{Turbo encoder.}
         \label{fig:turbo_enc}
     \end{subfigure}
     \\ \vspace{5mm}
     \begin{subfigure}[b]{1.0\linewidth}
         \centering
         \includegraphics[width=0.95\linewidth]{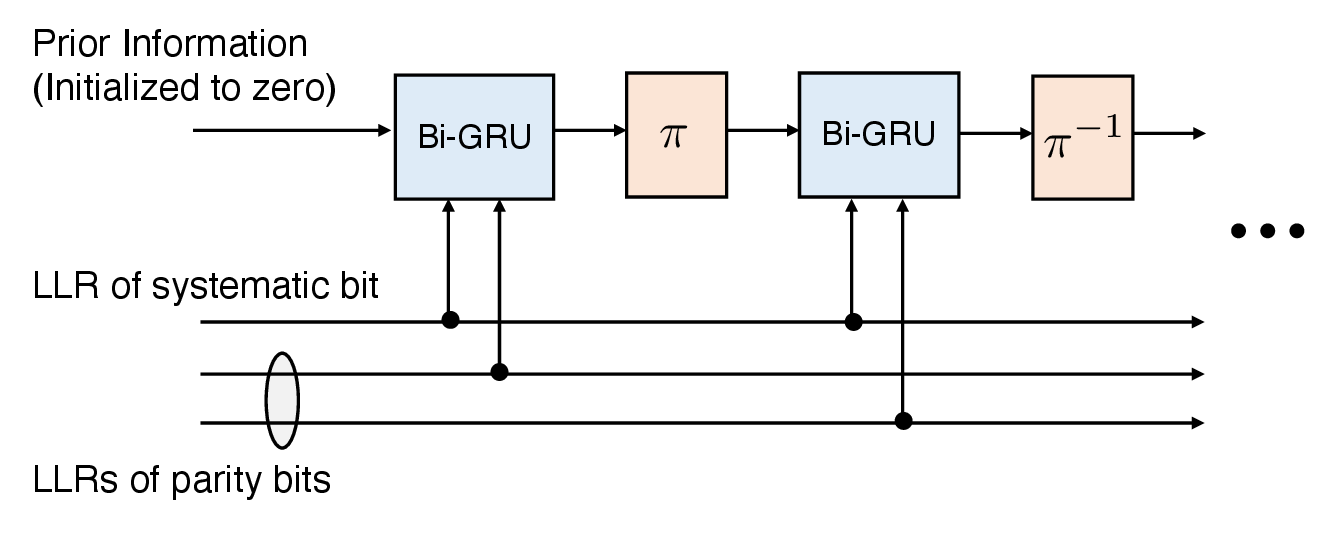}
         \caption{Model-free DL turbo decoders based on Bi-GRUs.}
         \label{fig:turbo_dec}
     \end{subfigure}
    \caption{An encoder and DL-based decoder of turbo codes. In the NEURALBCJR decoder \cite{kim2018communication}, each Bi-GRU is pre-trained to imitate BCJR algorithm, followed by an end-to-end training. In the DEEPTURBO decoder \cite{jiang2019deepturbo}, on the other hand, all Bi-GRUs are trained directly to optimize the end-to-end performance.}
    \label{fig:turbo}
\end{figure}

The aforementioned model-free turbo decoders, i.e., NEURALBCJR \cite{kim2018communication} and DEEPTURBO \cite{jiang2019deepturbo}, are illustrated in Fig.~\ref{fig:turbo}, where the constituent SISO decoders of the turbo decoder are replaced with Bi-GRUs without considering the specific trellis structure of the RSC encoders.
In contrast, the authors in \cite{he2020dnn} proposed a novel model-driven decoder architecture, called TurboNet, which integrates DNN into the traditional max-log-MAP algorithm. Furthermore, they applied network pruning to TurboNet to effectively reduce the number of parameters. The resulting TurboNet+ decoder was shown to achieve state-of-the-art performance and outperform existing DL turbo decoders even with lower computational complexity.

Subsequently, the authors in \cite{hebbar2022tinyturbo} proposed TINYTURBO which significantly reduces the trainable parameters of TurboNet+ by sharing the same weight across bit indices in the computation of the posterior LLR. In particular, for a block length of $40$, it was demonstrated that TINYTURBO with $18$ parameters outperforms TurboNet+ with $720$ parameters over AWGN channels. Furthermore, the strong adaptability of TINYTURBO to other block lengths, rates, and trellises, as well as its robustness to channel variations, were demonstrated.

\subsection{DL-Aided Decoding of Cyclic Codes}
\label{sec:CyclicDecoder}
DL decoders exploiting algebraic properties of cyclic codes have also been studied. In \cite{chen2021cyclically}, the authors proposed a neural network-based decoder for cyclic codes by exploiting their cyclically invariant property. More specifically, inspired by the fact that the maximum-likelihood decoder of any cyclic code is equivariant with respect to cyclic shifts, they imposed a shift-invariant structure on the weights of the neural decoder so that any cyclic shift of inputs results in the same cyclic shift of the outputs. Simulations of BCH codes and punctured \ac{rm} codes showed that the proposed decoder consistently outperforms the neural BP decoder proposed in \cite{nachmani2018deep}. Furthermore, they proposed a list decoding procedure that can significantly reduce the decoding error for BCH codes and punctured RM codes.

While the list decoding significantly improves the BLER, the major drawback was its relatively high BER. To improve the BER, the same authors proposed the improved version of the list decoder in \cite{chen2022improving}. The new decoder achieved a significantly lower BER compared to the list decoder in \cite{chen2021cyclically} while maintaining the same BLER.

\section{Conclusion}
\label{sec:conclusion}
In this paper, we have provided a comprehensive survey on DL for the channel coding problems. In particular, we have focused on DL methods for the code design and channel decoding problems. In what follows, we summarize the potential advantages and challenges of these approaches.

\subsection{Code Design Applications}
The conventional code design algorithms such as EXIT chart and variants of DE require ideal assumptions about channel models and decoding schemes. In contrast, the major advantage of using data-driven DL for code design is that one can tailor codes to more realistic channels and decoding schemes, for which theoretical analysis is intractable.

In Section~\ref{sec:design}, we saw that RL is a particularly popular approach to designing polar codes, among others. In this method, an agent learns to choose a new information bit position that minimizes the cumulative reward, which corresponds to the actual performance in terms of BLER. Calculating the reward requires Monte-Carlo simulations, which can be computationally intensive depending on the code length and target error rate. This complexity issue can hinder its application to scenarios with long code lengths and low error rates. Therefore, a design objective that can be efficiently computed during the training process may be desirable.

\subsection{Channel Decoding Applications}
As we have seen, channel decoding is a popular application of DL and a significant number of papers on this topic have become available. These approaches can be broadly classified into model-free and model-based approaches.

\subsubsection{Model-Free Approaches}
Model-free decoders employing a ``black-box'' neural network have the potential to outperform conventional decoding algorithms in terms of error rate performance and decoding complexity/latency. In particular, it has been demonstrated that model-free decoders can outperform existing decoding algorithms for short code lengths with highly parallelizable structures. This approach is thus potentially suitable for low-latency applications requiring short code lengths. Note that the performance is highly dependent on the DL model employed, and currently, the Transformer-based decoder achieves the state-of-the-art performance \cite{choukroun2022error}. However, the performance could be potentially improved by the advanced DL techniques.

Due to the curse of dimensionality, the applications of model-free decoders have generally been limited to short codes. In general, a larger model size is required to decode a longer code, which not only entails high computational complexity, but also high space complexity in both training and inference phases. Another concern about this method is the robustness against adversarial attacks \cite{madry2017towards}, as wireless networks are always vulnerable to radio jamming attacks \cite{pirayesh2022jamming,adesina2022adversarial} due to the openness of wireless channels. In particular, DL-based communication systems may have a higher risk of being disrupted by jamming attacks than classical systems \cite{sadeghi2019physical,chen2024air}.

Instead of completely replacing conventional decoders, employing a DNN to augment existing decoders is effective for arbitrary code lengths.
For example, model-free training has been extensively studied for DNN-based selection of flipping bit indices in SCF decoding as we reviewed in Section~\ref{sec:PolarDecoder}. 

\subsubsection{Model-Based Approaches}
In contrast to the model-free approach, the model-based approach realizes a scalable decoder by taking advantage of the knowledge of code structures and conventional decoding algorithms.
One of the most promising approaches is deep unfolding, which unfolds an iterative algorithm \cite{hershey2014unfolding} and introduces a set of trainable parameters. In particular, as we reviewed in Section~\ref{sec:BP}, neural network-based BP decoding over an unfolded Tanner graph augmented with trainable parameters \cite{nachmani2016learning} has been extensively investigated. As the underlying BP decoding algorithm has already been adopted in a wide range of communication systems, this approach can be applied to these systems with much less modification compared to model-free decoders. Many existing works have demonstrated that, by introducing and optimizing trainable weights that mitigate the effect of short cycles in the Tanner graph, the DL BP decoder can achieve better a trade-off between decoding performance and latency, i.e., the number of decoding iterations, compared to the standard BP decoder. This means that the performance advantage of the DL BP decoder becomes more significant as the number of short cycles increases.

\subsection{Challenges and Future Directions}
Despite their excellent performance, DL-based channel coding schemes face challenges that need to be addressed. We conclude this survey by highlighting several future research directions in this regard.

\subsubsection{Flexibility to Support Diverse Applications}
Next generation communication systems such as 6G will support heterogeneous applications that employ the channel codes with various block lengths, reliability, and latency requirements \cite{zhang2023channel}. Since there is no one-size-fits-all channel coding scheme, multiple code parameters, i.e., rates and lengths, must be supported to meet these requirements. On the other hand, adapting a DL decoder to different channels and code parameters would require an enormous amount of different parameter sets. This issue could be alleviated, for example, by a parameter sharing scheme and scalable GNNs as discussed in Section~\ref{sec:BP}. Furthermore, in order to support multiple code parameters, training must be performed multiple times, which is time consuming and computationally intensive. This complexity issue can be addressed by techniques such as transfer learning, meta-learning, and foundation models \cite{bommasani2021opportunities}.

\subsubsection{Explainable AI}
One of the disadvantages of DL methods is their black-box nature, which can hinder physical insights into the phenomena. Thus, evaluating and enhancing the explainability of generic DL models, i.e., the ability to provide reasons for the outcomes of the system \cite{phillips2020four}, remains an active field of research \cite{hoffman2018metrics,adadi2018peeking,xu2019explainable,linardatos2020explainable,arrieta2020explainable,minh2022explainable,dwivedi2023explainable}. In general, the explainability of DL models tends to have an inverse relationship to their performance, e.g., prediction accuracy \cite{gunning2019darpa}. Thus, a recent advanced DL model with a large number of parameters is particularly difficult to interpret and explain. 

In the next generation communications such as 6G, the concept of \ac{xai} will become increasingly important especially for the emerging mission-critical services, such as autonomous driving and remote surgery \cite{guo2020explainable,wang2021applications,wang2024explainable}. Although the effectiveness of DL for the physical layer has been demonstrated in terms of its performance, its explainability has not been well studied. Thus, XAI-based channel coding that increases the transparency of DL models and explains the reasons for decisions will be of practical importance. The new insights gained from XAI will also help us to devise code design and decoding algorithms.
    
\subsubsection{Efficient Training and Inference}
Recent advances in DL technologies have been driven by the exponential growth of data and computational power, with a focus on performance rather than the economic and environmental costs. This research trend is often referred to as \emph{Red AI} \cite{schwartz2020green}. Indeed, the computations required for DL algorithms result in a surprisingly large carbon footprint \cite{strubell2019energy,patterson2022carbon,wu2022sustainable}.
In contrast to Red AI, which prioritizes achieving state-of-the-art results, \emph{Green AI} aims to produce innovative results while taking into account computational costs \cite{schwartz2020green,salehi2023data}. This paradigm shift toward energy and cost efficiency is inevitable for the long-term success, and it is therefore important to carefully select and preprocess data, reduce redundancy, and avoid overfitting so as to minimize the amounts of data and computational resources required \cite{menghani2023efficient}.

Data-centric approaches are promising for reducing the energy consumption of DL algorithms \cite{salehi2023data}, while their primary goal was to improve performance in terms of accuracy\footnote{\url{https://datacentricai.org/}}. These approaches recognize that the quality of the training data has a significant impact on their performance, and thus prioritize data quality over model refinement \cite{jarrahi2022principles,zha2023data,zha2023survey}. These include active learning, knowledge transfer, dataset distillation, data augmentation, and curriculum learning. For channel decoding applications, some papers have employed these techniques for enhancing error rate performance, but more emphasis should be placed on the data efficiency.

\subsubsection{Quantum Machine Learning}
Quantum computing has a great potential to solve the classical channel coding problems \cite{botsinis2018quantum,matsumine2019channel,cui2022quantum,yukiyoshi2022quantum,kim2021heuristic,kasi2023quantum} more efficiently than digital computers.
Especially in the current \ac{nisq} era \cite{preskill2018quantum}, where the fidelity of quantum gates is limited by noise and decoherence, hybrid quantum-classical algorithms such as \ac{vqe} \cite{peruzzo2014variational} and \ac{qaoa} \cite{farhi2014quantum} are promising \cite{bharti2022noisy,cerezo2021variational}.
The potential of these algorithms for the classical channel decoding problems has been demonstrated in \cite{matsumine2019channel}.

Furthermore, \ac{qml}, which integrates quantum algorithms into ML, has received increasing attention \cite{schuld2015introduction,perdomo2018opportunities,carleo2019machine,nawaz2019quantum,huang2021power,simeone2022introduction,narottama2023quantum}.
For example, it has been shown that well-designed quantum neural networks can achieve a higher capacity and faster training ability than comparable classical feedforward neural networks \cite{abbas2021power}. Furthermore, quantum counterparts to classical CNNs, autoencoders, and \acp{gan} have been studied in \cite{henderson2020quanvolutional,romero2017quantum,dallaire2018quantum,lloyd2018quantum}. These methods have the potential to improve existing methods based on classical computers for a wide range of communication problems including channel coding.

\renewcommand{\IEEEiedlistdecl}{\IEEEsetlabelwidth{STT-MRAM}}
\MFUhyphentrue
\acsetup{
  uppercase/list,
  list/uppercase/cmd=\ecapitalisewords
}
\printacronyms[name=List of Abbreviations]
\renewcommand{\IEEEiedlistdecl}{\relax} 

\bibliographystyle{IEEEtran}
\bibliography{IEEEabrv,matsumine}

\end{document}